\documentclass[fleqn,usenatbib]{mnras}

\usepackage{newtxtext,newtxmath}
\usepackage{xcolor}
\usepackage{lineno}
\usepackage{tabularx}

\usepackage[T1]{fontenc}

\DeclareRobustCommand{\VAN}[3]{#2}
\let\VANthebibliography\thebibliography
\def\thebibliography{\DeclareRobustCommand{\VAN}[3]{##3}\VANthebibliography}

\usepackage{graphicx}	
\usepackage{amsmath}	

\title[EBL Constraints from Gamma-Ray Observations]{Revisiting EBL Constraints from Gamma-Ray Observations: A Critical Assessment and Methodological Improvements}

\author[K.~Abe~et.~al.]{\parbox{\textwidth}{\Large{
K.~Abe$^{1}$,
S.~Abe$^{2}$,
J.~Abhir$^{3}$,
A.~Abhishek$^{4}$,
V.~A.~Acciari$^{5}$,
A.~Aguasca-Cabot$^{6}$,
I.~Agudo$^{7}$,
I.~Albanese$^{8}$,
T.~Aniello$^{9}$,
S.~Ansoldi$^{10,38}$
% ...
}}\\
\\
\textit{Affiliations are listed at the end of the paper}\\
}
\date{Accepted XXX. Received YYY; in original form ZZZ}

\pubyear{2026}

\begin{document}
\label{firstpage}
\pagerange{\pageref{firstpage}--\pageref{lastpage}}
\maketitle
\vspace*{\fill}
\hypertarget{corrauth}{}%
\noindent\footnotesize $\star$ E-mail: \url{contact.magic@mpp.mpg.de}\par  
\clearpage
\begin{abstract} 
The extragalactic background light (EBL), ranging from the infrared to the ultraviolet bands, is the second most intense photon field in the universe, surpassed only by the cosmic microwave background (CMB). It is primarily generated by starlight in galaxies, either directly or through absorption by dust and re-emission at longer wavelengths. Very high energy (VHE, $E$ > $100\,$GeV) photons can be absorbed via $\gamma \gamma$ interactions with the EBL during their propagation across cosmological distances, providing an indirect method to probe the EBL by studying its impact on the energy spectra of distant VHE sources. This paper examines the robustness of EBL constraints derived from gamma-ray data, critically assessing the assumptions made in previous studies about the intrinsic source spectra, the uncertainties of the observations and the validity of the analysis tools. We find that earlier studies likely underestimated the uncertainties in the EBL intensity constraints, underscoring the need to account for systematic uncertainties comprehensively. By employing a Monte Carlo (MC) simulation and a plausible model for unknown systematic errors, we compute more realistic uncertainties. Additionally, we discuss possible alternatives to set EBL constraints, relaxing the assumptions on the intrinsic spectra of gamma-ray sources, with the goal of achieving more robust constraints.
\vspace*{10pt}

\end{abstract}




\section{Introduction}
The Extragalactic Background Light (EBL) is the second most intense diffuse photon field after the Cosmic Microwave Background \citep{Hauser_2001, Dwek_2013, Cooray_2016}. It is mainly composed of the accumulated light emission from stars in galaxies and other celestial objects such as active galactic nuclei (AGNs) over the cosmic history of the universe and redshifted by the expansion of the universe. Part of this light is absorbed by interstellar dust and re-emitted at longer wavelengths, which gives EBL its characteristic double-peaked spectral energy distribution.

There are different strategies for building EBL models and following \cite{dominguez_2011}, they can be divided into four different classes. The first two are forward evolution models using semi-analytical models of galaxy formation \citep{Gilmore_2012, Somerville_2012, Inoue_2013}, backward evolution models using extrapolations from local or low-redshift galaxy data to higher redshifts by making assumptions on galaxy evolution \citep{Franceschini_2008, Franceschini_2017}. The third class is based on inferred evolution from the cosmic star-forming history of the universe \citep{Kneiske_2004, Finke_2010, Khaire_2015, Andrews_2018, Finke_2022}, and the last class relies on observed evolution using galaxy data over a broad range of redshift \citep{dominguez_2011, Saldana_Lopez_2021}.

One EBL estimation method is based on counting photons from sources at cosmological distances in different photometric bands using deep galaxy survey data \citep{Pozzetti_2001, Keenan_2010, Tsumura_2013, Driver_2016}. 
This approach, generally known as "galaxy counts", can miss the contribution from faint, non-detected galaxies or the fainter outer regions of detected galaxies. Additionally, cosmic variance can introduce uncertainties in the measurements. Therefore, results obtained with this approach are usually treated as lower limits. Some works try to correct this via extrapolations of integrated galaxy light; this is what, e.g., \cite{Driver_2016} does, and this framework was recently updated by \cite{Koushan_2021} using deeper multi-wavelength survey data to provide more stringent constraints on the Integrated Galaxy Light (IGL) and significantly reduce measurement uncertainties.

Direct measurement of the EBL intensity using absolute photometry is possible, but challenging due to strong foregrounds, mainly zodiacal light and the brightness of our Galaxy \citep{matsumoto05, bernstein07, Matsuoka_2011}. \cite{Mattila_2017} use the shadowing effect of a high-galactic-latitude dark cloud on the EBL to estimate its intensity, a method recently improved by \cite{Haikala_2026} to provide a more up-to-date near-infrared measurement at 8600 \AA{} using the dark cloud DC303. Using measurements performed by the New Horizons probe beyond Pluto's orbit, where interplanetary dust glow should be negligible, \cite{Lauer_2022} reported a $4\sigma$ excess in the optical band compared to the integrated light from galaxy counts, but a recent reanalysis of the data by the same team \cite{Postman2024} resulted in a considerably lower value, indicating that background galaxies are the dominant, and perhaps even the sole source of the Cosmic Optical Background (COB).

The EBL in this range of wavelength can also be studied using a third method based on gamma-ray observations performed with space-borne or ground-based facilities. Among the latter, Imaging Atmospheric Cherenkov Telescopes (IACTs), see e.g. \cite{2022Galax..10...21S}, stand out as the leading instruments in the very-high-energy band (VHE, $E$ > $100\,$GeV).
VHE photons can interact with EBL photons to produce electron-positron pairs. Therefore, some of those photons are absorbed, which can be seen as an energy- and distance-dependent attenuation of the source flux.

This produces absorption features in the observed spectra of extragalactic sources which can be used to measure the EBL intensity \citep{Biteau_2015, Hess_2017, Biasuzzi_2019, Abeysekara_2019, Acciari2019, Desai_2019}. This method relies on making assumptions about the intrinsic energy spectra of the sources and comparing them with the observed ones, giving us information about EBL and its evolution. By "intrinsic" we mean "unattenuated", the spectrum as it would be recorded at Earth in absence of any absorption, i.e. if there was no EBL. It is not the same as the spectrum at the source, due to the redshifting of the photon energies. The analysis usually consists in a Likelihood Ratio Test (LRT) in which the EBL is represented by one or more free parameters, and in which the parameter uncertainties are computed analytically assuming that Wilks' Theorem\footnote{Wilks' Theorem states that under certain regularity conditions, 2 times the difference in the log-likelihoods of two nested models follows a $\chi^2$ distribution. The regularity conditions are large sample size (assumes large-sample conditions) and parameters in the interior of the parameter space.} \citep{Wilks_1938} can be applied.  Recent work by \cite{greaux2024} addresses the problem using a Bayesian approach to derive posterior distributions of the EBL parameters - it must be noted, however, that some of the reasons which may make Wilks' theorem fail would also affect the reliability of the results using a Bayesian approach. If the uncertainties are underestimated, the likelihood results do not show the correct probability of the observations, and this affects both the frequentist and Bayesian approaches.

The other main assumption made in this method is that the intrinsic spectrum of a source can be well described by a simple concave function with 2 to 4 free parameters (sometimes over several orders of magnitude in energy, when data from space and ground-based telescopes are combined).

The goal of this paper is to revise the validity of these assumptions through the use of a Monte Carlo (MC) simulation (using MAGIC observations as a test sample) and to propose methods for improving the robustness of the EBL estimates. This is particularly important considering that anomalies in EBL measurements based on gamma-ray observations have often been interpreted in the literature as hints of the existence of axion-like particles (ALPs), either to explain a too-low gamma-ray opacity of the universe (via photon-ALP conversions, e.g. \cite{Sanchez_Conde_2009}, \cite{Kohri_2017}, and \citealt{galanti2024}), or a too-high one (through an extra EBL component from the decay of light ALPs into photons e.g. \citealt{Korochkin_2020}). We are confident that such claims would be significantly weakened (if not refuted), should all subtleties of the gamma-ray analysis be properly taken into account.

The MAGIC data used in this paper are the same list of events used in \citet{Acciari2019}. That paper additionally included lower energy data (from $0.1$ to $\simeq\!\!100\,$GeV) taken by the Large Area Telescope (LAT) on board the Fermi Gamma-ray Space Telescope during similar time ranges as the MAGIC observations. Given the scope of this work, we decided not to use the contemporaneous Fermi-LAT data: on the one hand, they are not strictly simultaneous with MAGIC observations, which, given the variability of the sources, is a potential source of systematic errors. On the other hand, imposing a simple functional form for the spectrum over three decades in energy is obviously a stronger assumption than using the same function on the narrower range covered by MAGIC. Additionally, systematic uncertainties due to differences in energy scale calibration between Fermi-LAT and MAGIC are non-negligible and difficult to quantify. Therefore it seems at present not possible to estimate the (source- and sample-dependent) systematic uncertainties that would result from the use of Fermi-LAT data in the analysis and therefore we conclude that restricting the analysis to MAGIC data provides a more controlled and interpretable assessment of EBL constraints.

The paper is organized as follows. In \autoref{sec:analysis} we introduce the methodology used for the analysis: a maximum likelihood method identical to the one used in \cite{Acciari2019}, but re-implemented in Python, and a Monte Carlo simulation that we use to assess the validity of the procedure (in terms of goodness-of-fit and coverage). In \autoref{sec:MBPWL} we address the critical role of the intrinsic source spectrum modeling, discussing how standard assumptions affect results and presenting possible alternative approaches based strictly on a concavity constraint. In \autoref{sec:results} we present the results of this evaluation and propose ways to address the problems that were noticed with the reliability of the results. A sample of EBL-constraining MAGIC observations is used to show the effect of the proposed modifications of the method. Finally, \autoref{sec:conclusions} presents a summary and conclusions of the paper.

\section{Data Analysis}\label{sec:analysis}

\subsection{Maximum Likelihood Method}\label{sec:logL}
In this analysis, we adopted a maximum likelihood approach similar to the one in \cite{Abdo_2010}, \cite{Ackermann_2012}, \cite{Abramowski_2013}, \cite{Abdollahi_2018} and   \cite{Acciari2019}.
We will refer to this method as \textit{the classical approach} in the following. The method is implemented in an open-source Python software created for this analysis.\footnote{The code can be found in: \url{https://github.com/R-Grau/EBLpy}} \\
Since the data in the different bins of reconstructed energy are statistically independent, the likelihood function to maximize is the product of the likelihoods for all the bins in which we have data:
\begin{equation}
    L(ebl, \theta) = \prod_{i=1}^{N_{bins}} L_i(ebl, \theta).
    \label{eq:likelihood_total}
\end{equation}
Here $ebl$ represents the parameter(s) describing the EBL, while $\theta$ is the vector of the parameters that describe the intrinsic source spectrum.

When dealing with more than one spectrum (e.g. spectra from different sources or from the same source but in a different state), the joint likelihood is simply the product of the likelihoods $L$ for each individual spectrum. \\
With $N_{\text{on},i}$ and $N_{\text{off},i}$ being the recorded numbers of events in bin $i$ (after all analysis cuts) in the ON- and OFF-source regions, the $L_i$ term of eq. (\ref{eq:likelihood_total}) has the form:

\begin{align}
\begin{split}
    L_{i}(ebl, \theta) = \,\,&\textrm{Poisson}(g'_{i}(ebl, \theta) +b_{i}, N_{\textrm{on},i})\,\cdot \\
    &\textrm{Poisson}(b_{i}/\beta, N_{\textrm{off},i})\cdot \textrm{Gauss}(g'_{i},g_{i},\Delta g_{i}),
    \label{eq:likelihood}
\end{split}
\end{align}
where the $b_i$ are the nuisance parameters for the Poissonian background that is recorded alongside the gamma-ray signal in the ON region. $\beta$ is the ratio of the ON to OFF exposure. $g'_{i}$ are the Poisson parameters of the signal (mean number of gamma rays in the ON-source region for bin $i$).\\
The $g'_{i}$ are {\it not} completely determined by folding the tentative intrinsic energy spectrum with the EBL absorption and the instrument response function (IRF). The reason is that we incorporate in the likelihood the statistical uncertainty of the IRF (computed with the MAGIC MC simulations): we assume $g'_{i}$ follows a Gaussian of mean $g_{i}$ and standard deviation $\Delta g_{i}$. Both values are obtained by folding the absorbed spectrum with the IRF, propagating the IRF uncertainties, and multiplying the resulting rates by the observation time. Then a Gaussian term is added in the likelihood and the $g'_{i}$ are treated as nuisance.\\
In each step of the numerical likelihood maximization process (in the space of the $ebl$ and $\theta$ parameters) the values of $b_i$ and $g'_{i}$ which maximize $L_i$ can all be determined analytically.

As usual, instead of maximizing the likelihood, we will minimize the quantity $-2\textrm{log}(L)$ using minuit.\\
We define $L_{max}$ as the maximum likelihood achieved in the ($ebl$, $\theta$) free parameter space, and $L^*$ as the maximum (unconstrained) value of the likelihood (of a model that predicts exactly the observed values of $N_{\text{on},i}$ and $N_{\text{off},i}$ in every bin of reconstructed energy). Let $m$ be the number of free parameters for the intrinsic spectral model $\theta$, and $n$ the number of free parameters to describe the EBL. Then, if Wilks' theorem can be applied, and the model is a good representation of the data, the quantity $-2\textrm{log}(L_{max}/L^*)$ is distributed as a $\chi^2$ with $N_{bins} - n - m$  degrees of freedom. This allows us to compute the goodness of fit, i.e. the P-value of the best-fit.\\
On the other hand, we can perform a LRT of the model with free EBL parameters vs. a simpler model with fixed EBL (e.g. no EBL at all, or a fixed template model). With $L'_{max}$ being the maximum likelihood for the simpler model, Wilks' theorem states that $-2\textrm{log}(L'_{max}/L_{max})$ is distributed as a $\chi^2$ with n degrees of freedom. From the profile likelihood of the $n$ EBL parameters (considering all other as nuisance) around the minimum we can obtain the uncertainties (and correlations) of the $n$ EBL parameters. 
In the rest of this paper we consider the case $n=1$, i.e. a single EBL parameter, denoted by $\alpha$. The parameter $\alpha$ is a dimensionless global scaling factor that rescales the optical depth predicted by a reference EBL model ($\alpha = 1$ corresponding to the unmodified template); for a fixed template, it is therefore also referred to as EBL intensity scale.


In order to validate the Python code developed for this study, we applied it to the analysis of several MAGIC datasets: the February 2014 flare of 1ES1011+496 (average spectrum), at $z = 0.212$,  and 15 different spectra of Mrk421 ($z = 0.031$) in high state obtained in 2013 and 2014. We compared the results with those reported in \cite{Acciari2019}, which were obtained (from the same datasets) using the ROOT-based \citep{BRUN199781} MARS software package \citep{moralejo2009mars}, the official MAGIC analysis software. Specifically, we compared the profile likelihood of $\alpha$ relative to the \cite {dominguez_2011} model, in the range 0 to 2, and only negligible differences (with no impact on the best-fit values of $\alpha$ and their uncertainties) were found.
    
\subsection{Monte Carlo simulation}\label{sec:ToyMC}
The validity of Wilks' theorem, which holds in the asymptotic limit of very large event statistics, cannot be guaranteed in practice, and it is therefore advisable to study each specific problem with the help of Monte Carlo methods. Besides the finite event statistics, the use of models with limits in the range of their free parameters (in our case, the requirement that the intrinsic source spectra are concave - see \autoref{sec:MBPWL}) may also lead to the non-fulfillment of the theorem. A Monte Carlo can be used to determine if Wilks theorem can be applied to the case of EBL determination; in case it is not applicable, the simulation can provide the distribution of the likelihood ratio, and hence the uncertainty of the parameter of interest for the desired coverage.

The Monte Carlo developed for this work consists of two parts. The first one simulates an observation of a flaring source. The second one uses this simulated data to make a profile likelihood of the scaling factor of the EBL density ($\alpha$) and find its most likely value. The procedure is repeated for multiple Poissonian realizations of the data, to study the distribution of the results.

For the simulation of the observation of a flaring source, we first select the function to model the intrinsic differential energy spectrum ($dF/dE$) of the emission. The options that are available in our code are the power law (PWL) with 2 parameters; the log parabola (LP) and the power law with exponential cut-off (EPWL), with 3 parameters; and the log parabola with exponential cut-off (ELP) and the power law with super-exponential cut-off (SEPWL), with 4 parameters. We have chosen these functions because they are commonly used to describe SEDs in the VHE range. The parametrizations of the functions are the following:

\begin{alignat*}{2}\label{eq:PWL}
& \text{PWL:} &\quad & F_0  (E/E_0) ^{-\Gamma},\\
& \text{LP:}  && F_0  (E/E_0) ^{-\Gamma-b\: \log_{10}(E/E_0)},\\
& \text{EPWL:} && F_0  (E/E_0) ^{-\Gamma} \: e^{-E/E_c},\\
& \text{ELP:}  && F_0  (E/E_0) ^{-\Gamma-b\: \log_{10}(E/E_0)} \: e^{-E/E_c},\\
& \text{SEPWL:} && F_0  (E/E_0) ^{-\Gamma} \: e^{-(E/E_c)^d},\\
\end{alignat*}
where $E_0$ is the normalization energy and $F_0$, $\Gamma$, $b$, $E_c$ and $d$ are parameters.

Once the function and parameter values for the simulation have been selected, the effect of the EBL is introduced using the energy-dependent optical depth taken from a given EBL model. The resulting absorbed differential energy spectrum is folded with the MAGIC instrument response function, using an energy migration matrix that incorporates the exposure: (effective area $\times$ observation live time) vs. reconstructed energy ($E_\text{reco}$) vs. true energy ($E_\text{true}$). The result of this folding is the expected number of signal events in each bin of reconstructed energy. Adding the expected number of background events for the observation time, we obtain the expected number of ON-source events. The expected background is scaled by $1/\beta$ (see eq. \ref{eq:likelihood}) to obtain the expected number of events in the OFF-region. The actual number of ON- and OFF-source events in each realization are generated following Poisson distributions with the respective expected values. 
The maximum likelihood analysis of each realization is performed following the procedure explained in \ref{sec:logL}.

To simulate a MAGIC observation, we use the nominal intensity of the chosen EBL model (i.e., we set the scaling parameter $\alpha$ to 1) and use the best-fit intrinsic spectrum obtained from the actual data under this assumption. 
For the analysis, we leave $\alpha$ free, and we model the intrinsic spectrum with the same function (with free parameters) used to simulate the observations or with a different one, to test the effect of assuming in the analysis a function that does not exactly reproduce the true spectral shape.

\subsection{Systematic uncertainties}

By simulating the MAGIC observations mentioned at the end of section \ref{sec:logL} (and analyzing them with the "correct" intrinsic spectral shape assumption), we noticed that, while the shape of the profile likelihoods was similar to the one obtained with the real data, the P-values of the latter were much lower. This was not surprising: as already noted in \cite{Acciari2019} the P-values of the best-fit models reported there for the same datasets are rather low, O($10^{-2}$). The low quality of the fits might originate in any of the ingredients of the model, i.e. in differences between our assumptions and reality in the intrinsic spectral shape, or in the EBL model\footnote{With $\alpha$ as the sole free EBL parameter, the evolution and spectral shape of the EBL are fixed inside $\tau (E,z)$ as provided by the model, and could be different from the real ones. No EBL model uncertainties are considered in this type of analysis.}. However, the fit residuals do not show any smooth pattern vs. energy which suggests that the problem could be solved with a more complex model. Another possibility is that the problem is on the data side, specifically in the possible presence in the flux measurements of hidden systematic errors which are non-negligible compared to the statistical uncertainties considered in the analysis. Such instrumental systematics are not unexpected: we compute the IRFs from a detailed MC simulation of the telescope and of the shower development in the atmosphere, but there are unavoidable simplifications that can result in systematic differences between the real and the simulated performance. Such systematic errors can be energy-dependent and time-dependent. Previous works often take into account a possible overall mismatch of the "energy scale" of real and simulated telescopes (arising for instance from variations in atmospheric transparency), and evaluate its effect on the EBL determination by shifting the reconstructed spectra along the energy axis, or by changing the global light collection efficiency in the simulation. This, however, does not address the issue of the too low P-values.

In order to reproduce this effect in the Monte Carlo, we adopted the following model of energy-dependent systematics: a Gaussian-distributed error in the effective area (i.e. a difference between the area used for the Monte Carlo generation and the one assumed in the analysis), independent in each energy bin, and with the same relative standard deviation (a given fraction of the effective area) through the whole energy range. This is a simple way of simulating these systematics, which assumes no correlations between the systematics in the different energy bins, but, of course, there are other ways of doing it. Our goal is to test how big the effect of the hidden systematics (probably responsible for the low P-values) can be in the EBL constraints. Future telescopes are expected to have improved calibration methods, which will allow the estimation of these systematic uncertainties. In section \ref{sec:results}, we will tune this relative systematic error so that we achieve a good match between the profile likelihood from the real data analysis and the median profile from the MC simulation realizations, and test the impact on the coverage of the EBL constraints.

Besides introducing these systematic errors in the MC simulations, one can also incorporate them as systematic {\it uncertainties} in the analysis, by adding them to the likelihood in eq. \ref{eq:likelihood} in the Gaussian term (which originally accounts only for the IRF statistical uncertainties due to finite MC simulation statistics). This is easily done by increasing the $\Delta g_i$ by a fraction of $g_i$. In that way, we would allow the $g'_i$ nuisance parameters to account for these systematics, hence obtaining more reasonable P-values. This naturally results in a flatter profile likelihood for $\alpha$, which in turn affects the computed uncertainties, as we discuss in section \ref{sec:results}.

We must stress that the use of P-values to assess the quality of the fits, and the possible presence of hidden systematics, requires that the analysis makes use of the instrument response function (energy migration and effective area) of the telescopes, like e.g. in the "forward-folding" approach used by MAGIC. Analyses based on published data points (flux vs. true energy), which have no access to (and hence disregard) the point-to-point correlations, will typically display overestimated P-values which could mask the effect of hidden systematic errors.

\section{Modelling the intrinsic source spectrum}\label{sec:MBPWL}

The most critical assumption needed to set EBL constraints using gamma-ray observations is the one about the intrinsic spectral {\it shape} of the high-energy emission. Typically, the differential energy spectrum is assumed to be well described by a simple function with two to four free parameters, like those presented in section \ref{sec:ToyMC}. Besides, we assume that the function must be concave anywhere in the fitting range, i.e. the spectrum cannot become harder as energy increases. Having several possible functions to choose from is problematic; strictly speaking, for the results of the Likelihood-ratio test (namely, the estimate of $\alpha$) to be reliable, one should be certain that the chosen function is an accurate model of the intrinsic spectrum. In spite of this, the choice among the candidate functions is usually made based on the quality of the resulting best fits, either via the P-value, as in \cite{Acciari2019}, or using a LRT. In the latter case, one can require a given model to improve the fit quality significantly (e.g. by $2\sigma$ like in \citealt{Biteau_2015}) over (nested) models with fewer parameters\footnote{This may be dangerous, though: one may choose the simplest model, a power-law, for many spectra which individually have <$2\sigma$ evidence for curvature. Used together in a joint analysis, the small intrinsic curvature of each spectrum may end up being {\it attributed} to the absorption, and hence produce a bias towards too high $\alpha$.}. 

As we show in the following (see \autoref{subsec:1ES}), EBL constraints can be highly dependent on the choice of the fitting function. This dependency is also shown in \cite{Biasuzzi_2019} Figure 1. More generally, the basic problem is that the {\it overall} imprint of the EBL on the gamma-ray spectrum of a source, namely a softening of the spectrum and a cut-off, cannot be unambiguously attributed to the absorption process: both are features that are potentially present already at the source. Ideally, the functions used to model the intrinsic spectrum should be able to describe all its possible features (even if degenerate with EBL absorption features), otherwise we risk attributing them to the EBL, and hence derive biased EBL constraints.

A recent example that could illustrate this issue is provided by the results obtained by the LHAASO collaboration in \citep{LHAASO_2025} using GRB~221009A data \citep{LHAASO_2023}. Their analysis of six spectra (not fully independent) consistently shows larger absorption than predicted by commonly used EBL models, such as \cite{Saldana_Lopez_2021}. One possible explanation is that the intrinsic spectral function selection (based in the Akaike Information Criterion) is just choosing the simplest functions that can {\it model} the observations, and in doing so it may be interpreting source features as part of the absorption imprint: indeed, out of the six spectra, four are modeled with a power-law, hence any curvature observed in the data has to be attributed to EBL absorption, potentially biasing their results towards higher EBL density.

Fortunately, the EBL imprint has a characteristic feature at energies between a few 100 GeV and a few TeV: a "wiggle" (double inflection point) in the log-log representation of the transmissivity ($e^{-\tau}$) vs. energy. This feature is uniquely well-located within the optimal sensitivity window of IACTs and is particularly distinct for sources at redshifts from 0.1 to 0.5: at lower redshifts, the absorption is too subtle to distinguish from systematic errors , while at higher redshifts, the severe attenuation suppresses the source flux entirely in this band. This specific energy-redshift window allows us to distinguish the EBL signature from smooth intrinsic curvature more robustly. Such a feature should not be present in the intrinsic spectra, which are expected to be concave (i.e., softer as energy increases) in the energy range covered by MAGIC and other IACTs. This is consistent with observations, since none of the nearby VHE sources (unaffected by EBL) shows any significant convexity within the studied energy range \citep{Abdo_2010}. In addition no significantly convex spectrum is reported in the Fermi-LAT fourth source catalog \citep{Abdollahi_2020, Ajello_2020} for any source, including many AGNs for which LAT sees the Compton peak in the energy band immediately below the one covered by IACTs.

While recent studies such as \cite{Paliya_2025} have revealed that complex AGNs, particularly Flat Spectrum Radio Quasars (FSRQs), can show additional spectral components that induce convexities, these sources are less central to IACT studies due to their sharp flux cutoffs in the VHE regime. Moreover, those results rely on data integrated over many years, whereas IACTs typically detect sources in flaring states. For High Synchrotron Peaked (HSP) BL Lacs (the majority of blazars detected by IACTs), \cite{Dinesh_2025} show that such additional components are typically absent on the timescales of days to weeks characteristic of these events. Therefore, the assumption that the sources in our sample are simpler and dominated by a single emission component is physically well-motivated.

The question is whether we can obtain EBL constraints with the sole assumption that the intrinsic spectrum is concave, with no additional requirement on its shape. We have considered two possible approaches to achieve this, which are presented in the rest of this section.

\subsection{Multiply broken power-law}\label{subsec:mbpwl}

The first approach consists of using as a model for the intrinsic spectrum a "generic concave function" that would be able to fit any concave shape of the spectrum, but not the EBL-induced wiggle. If the imprint of the EBL is significant enough in the reconstructed energy spectrum, a certain minimum level of EBL will be needed to achieve a good fit, hence providing a lower bound to $\alpha$. On the other hand, a too high $\alpha$ could over-correct the wiggle, as well as the EBL cut-off, and call for a non-concave intrinsic spectrum - hence an upper bound is also attained.

To exemplify this approach, we opted for using the Multiply Broken Power-Law (MBPWL) as intrinsic spectral model: a Power-Law ($\propto E^{-\Gamma}$) defined by parts, with the photon index, $\Gamma$, changing at points known as nodes or breaks. The concavity constraint is satisfied by imposing $\Delta\Gamma>=0$ in every break.

With sufficient number of breaks, the MBPWL should fit quite well the spectrum of a source even under the assumption of no EBL ($\alpha = 0$), except for the wiggle if it is significant, due to the concavity constraint. For the same reason, it should struggle to fit the spectrum for large values of $\alpha$ (the value at which this happens depends on the curvature of the intrinsic spectrum of the source).
To define the MBPWL, we need the usual PWL parameters of the first segment ($F_0$ and $\Gamma$), the number of nodes, their corresponding energies, and the change in $\Gamma$ at each node. 
For a fixed number of nodes ($n$), the number of free parameters is $N_p = 2\,(n+1)$. As $n$ increases, the number of parameters increases rapidly, which can lead to convergence issues during the likelihood maximization process. 
These convergence issues typically arise when there is degeneracy among the parameters of the model. 
While degeneracy between (plausible) intrinsic spectral features and the imprint of EBL absorption is a real problem that should not be hidden away by reducing the flexibility of the spectral function, internal degeneracy among the parameters of the latter should indeed be avoided.
This makes the MBPWL a problematic function: even for a fixed number of breaks, leaving the break energies as free parameters often leads to fit convergence issues. For this reason, we do not think that the MBPWL can be adopted as a generic replacement of the simpler models so far used in the literature. Still, we will use it to show a specific example (the analysis of the 1ES1011+496 flare that we present in section \ref{sec:results}) in which the adoption of the MBPWL results in a major change of the lower constraint of the EBL intensity.  We fixed the number of nodes and their energies to those which best fit the observed spectrum (i.e. setting alpha=0), adding nodes one at a time and scanning their position (i.e. break energy). The number of nodes is increased as long as the change of $\Gamma$ in the latest added node is larger than $0.01$ \footnote{This is just a somewhat arbitrary safe condition to ensure that all nodes correspond to a non-negligible change in $\Gamma$.}. The idea of using $\alpha = 0$ in this process is to guarantee that the number of nodes is sufficient to account for the full curvature of the {\it observed} spectrum (except for the possible non-concavity induced by the EBL). Then, the MBPWL with fixed break energies (and normalization and photon indices as free parameters) is used as intrinsic spectral model in the determination of the EBL intensity. Any increase of the maximum likelihood as we increase $\alpha$ could be attributed to the reduction of the convex feature induced by the EBL. The resulting EBL constraint would be valid under the generic assumption that the intrinsic spectrum is concave (instead of e.g. "it is a concave log parabola"). Of course, the MBPWL, with its sharp changes of photon index at the nodes, is only an approximation of a truly general concave function - but certainly way more flexible than the few parametrizations presented in section \ref{sec:ToyMC}. For the MC simulation (see \autoref{subsec:simulation}), to reduce the computing time, we get the optimal number of nodes and their position with the representative Asimov dataset obtained by using the mean value of ON and OFF events instead of the Poisson realization, and fix those values for all the Poisson realizations.

\subsection{Focusing on the EBL inflection points: the "concave EBL" approach}\label{sec:concaveEBL}
The use of the MBPWL as an intrinsic spectral model to reduce the number of assumptions in the EBL determination is not a fully satisfactory solution. The number and positions of the nodes are not free in the likelihood maximization (to reduce the occurrence of convergence issues), but fixed in advance through a "recipe" that is somehow analogous to the procedures used in previous works to choose among different intrinsic spectral functions. This amounts to having "hidden" free parameters, which can compromise the soundness of the statistical approach. 

To overcome these issues, we have considered an alternative method to increase the robustness of the EBL constraints by making them less dependent on the assumptions about the intrinsic spectrum. The idea is to conservatively assume that the "concave part" of the $e^{-\tau}$ vs. $E$ curve, since it is concave, could be an intrinsic feature of the spectrum and not an imprint of the EBL. We will try to constrain $\alpha$ only through the deviations from the concavity that it imprints in the spectrum.

\begin{figure}
    \centering
    \includegraphics[width = \linewidth]{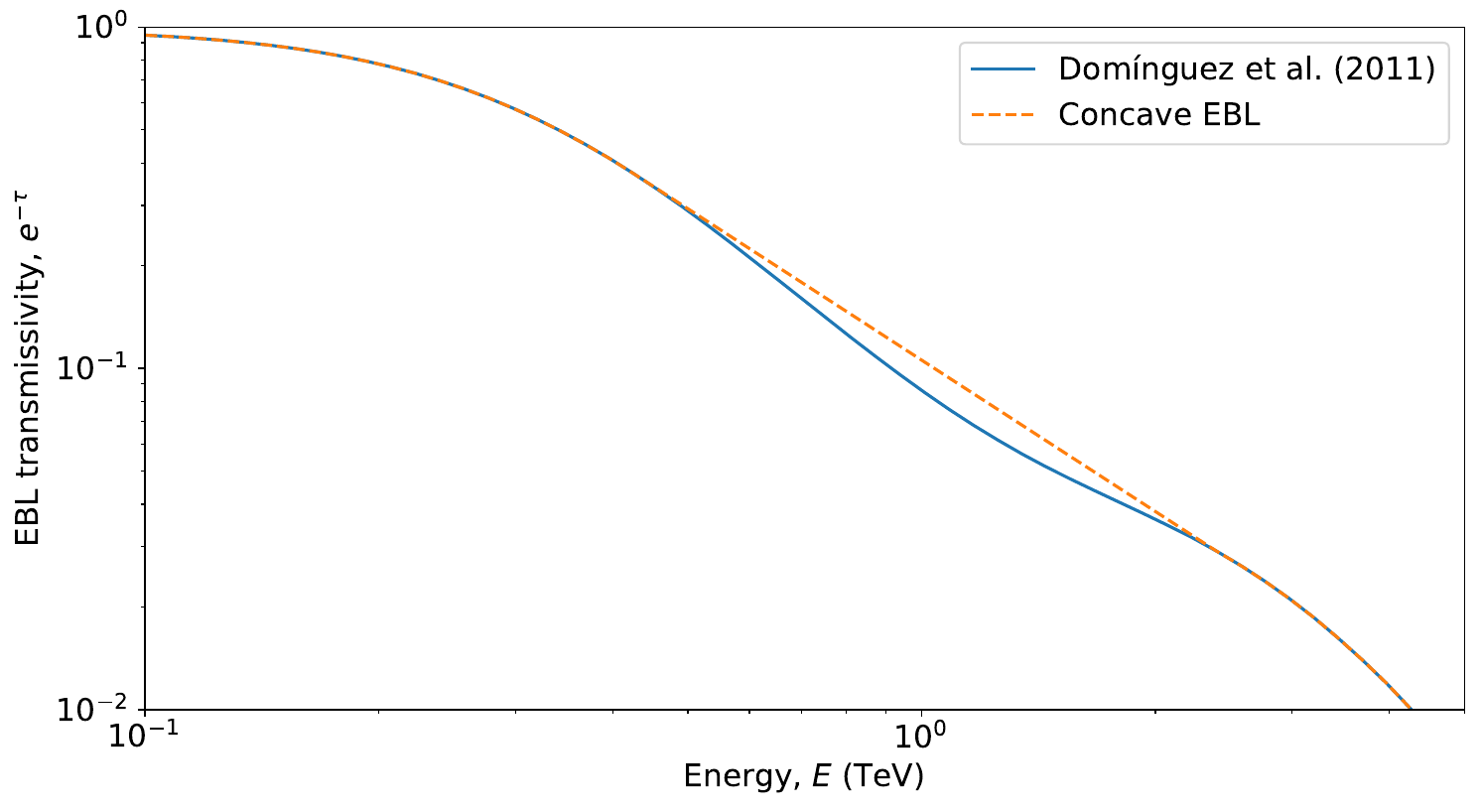}
    \caption{Blue solid line shows the EBL transmission factor $e^{-\tau}$ for $z = 0.212$ (the redshift of 1ES1011+496) compared to the dashed orange line which shows the modified EBL transmission factor ($e^{-\tau'}$) in which the inflection points (i.e. "the wiggle") have been removed.}
    \label{fig:EBL_concave_comp}
\end{figure}

We start by defining a concave version (in log-log representation) of the EBL transmission curve, $e^{-\tau'}$ vs. $E$. We derive it from the usual EBL absorption ($e^{-\tau}$), just replacing the convex part by a power-law tangent to the curve on both sides of the feature (see \autoref{fig:EBL_concave_comp}, where the result is compared to the original \cite{dominguez_2011} curve for a redshift $z = 0.212$). We choose a PWL tangent to the EBL absorption as it is the way to remove the inflection points which adds minimal concavity. We compute the profile likelihood using the following expression for the {\it absorbed} source spectrum:

\begin{equation}\label{eq:tauprime}
    \frac{dF}{dE}\bigg\rvert_{\text{obs}}(E) = \frac{dF}{dE}\bigg\rvert_{\text{int}}(E)\cdot e^{-\tau'(E,z)} \cdot e^{-\alpha(\tau(E,z) - \tau'(E,z))}.
\end{equation}

Here $dF/dE$ is the model for the intrinsic spectrum (a parametric function, just like in the usual approach), $\alpha$ is a dimensionless parameter that scales the convex feature, $\tau(E)$ is the optical depth predicted by the EBL model and $\tau'(E)$ is the modified optical depth. For $\alpha = 1$, the convex feature corresponds to that of the nominal EBL model, while $\alpha = 0$ implies no convexity. The physical meaning of $\alpha$ is therefore consistent with that adopted throughout this work, a global scaling of the EBL optical depth relative to the reference model.

In this method we still need to make some assumption on the shape of the intrinsic spectrum, $dF/dE$. For a given function $dF/dE$ (e.g. a log parabola) the difference with the usual approach (in which $\alpha$ scales the total optical depth) is that we will not obtain an artificially\footnote{In the sense that it is valid only under the strong assumption that the intrinsic spectrum is not just concave, but also follows the assumed function $dF/dE$.} good lower bound for $\alpha$ just because $dF/dE$ fails to fit the overall {\it concavity} of $e^{-\tau}$ vs. $E$. The increase of $-2\log L$ as we decrease $\alpha$ from its best-fit value will come exclusively from the inability of the (concave-defined) $dF/dE$ to model the EBL-induced convexity. Note that in \autoref{eq:tauprime} we can as well replace the nominal $\tau(E)$ in the EBL model by a scaled version that provides a better best-fit (obtained with the classical approach).

An intuitive way of visualizing this method is to imagine that we have a high-quality dataset (i.e. a precisely measured energy spectrum) like those we expect to obtain when observing AGN flares with the next generation of IACTs. The observed spectrum has a significant EBL-induced convex feature. Now, we correct the spectrum for the EBL absorption using a scaled version (by a factor $\alpha$) of our template EBL model. As we increase the value of alpha, we see the de-absorbed spectrum progressively lose its convex feature. At some point the spectrum will be statistically compatible with a concave shape (no matter which particular one). This would set a lower bound on $\alpha$ whose validity just relies on the assumption that the intrinsic spectrum is concave\footnote{Note however that {\it in our implementation} using \autoref{eq:tauprime} dF/dE must be a plausible model {\it for the best-fit $\alpha$} - to be assessed via the P-value.}. For large values of $\alpha$ that over-correct the convex feature, the spectrum will again become incompatible with the concavity constraint. This however does not provide a competitive upper bound for $\alpha$, since we do not de-absorb the EBL cut-off at the highest energies, and hence we do not induce the "pile-up" that makes $-2\log L$ grow fast  in the classical approach.

The proposed "concave EBL" method is just one way to implement this intuitive idea, one which allows to combine it with the forward-folding that we apply to account for the instrument response. The method could in principle be implemented in a different way, e.g. via a spectral unfolding which provides spectral flux points (vs. true energy) with their uncertainties {\it and} correlations, and some statistical test which evaluates for each $\alpha$ the significance of the deviation of the de-absorbed spectrum from concavity.

\section{Results}\label{sec:results}
\subsection{Monte Carlo study of the validity of the classical approach} \label{subsec:simulation}
We first use the MC simulation to evaluate the performance of the classical approach applied to the reference MAGIC datasets introduced at the end of \autoref{sec:logL}.

For each target spectrum, we simulate 10,000 Poisson realizations of the observation. The Poisson parameters for the background simulation are obtained from the Off-source counts in the real observation, normalized to the size of the On-source region, while for the signal we use the best-fit spectrum. All values are taken from the analysis presented in \cite{Acciari2019}:
\begin{itemize}
    \item For the simulation of 1ES1011+496 (February 2014 flare) we used a PWL with photon index $\Gamma = 2.03$ and normalization factor at 250 GeV $F_0 = 8.70 \cdot 10^{-6} \mathrm{m^{-2} s^{-1} TeV^{-1}}$, with the source located at $z=0.212$ and observed during $11.8\mathrm{h}$.
    \item   For the different Markarian 421 spectra, the functions and parameters used are reported in \autoref{tab:Mrk_funcs}, taken from the analysis in \citealt{Acciari2019} (although, being nuisance parameters, they were not reported in the article).
\end{itemize}

For the simulation of the EBL absorption we adopted the \cite{dominguez_2011} EBL model in order to compare the results with \cite{Acciari2019}. 

As explained at the end of \autoref{sec:ToyMC} we first have to look for the level of systematic errors that we need to add to the measured fluxes in order to reproduce the results obtained with the real observations, i.e. to achieve (on average) a similar best-fit P-value and profile likelihood curve for $\alpha$. We tested different values of standard deviation for both 1ES1011+496 and Mrk421. The standard deviations of the systematic errors that provide the best match are $8.5\%$ and $3.3\%$ for 1ES1011+496 and Mrk421 respectively.

The values for both sources are different as these systematics can be combination of different factors: systematic errors in the effective area (or more in general, in the instrument response function), discrepancies between the fit function and the source's true spectrum, and differences between the EBL model and the actual one (which may be redshift-dependent). 

\subsubsection{1ES1011+496}\label{subsec:1ES}
\begin{figure}
    \centering
    \includegraphics[width = \linewidth]{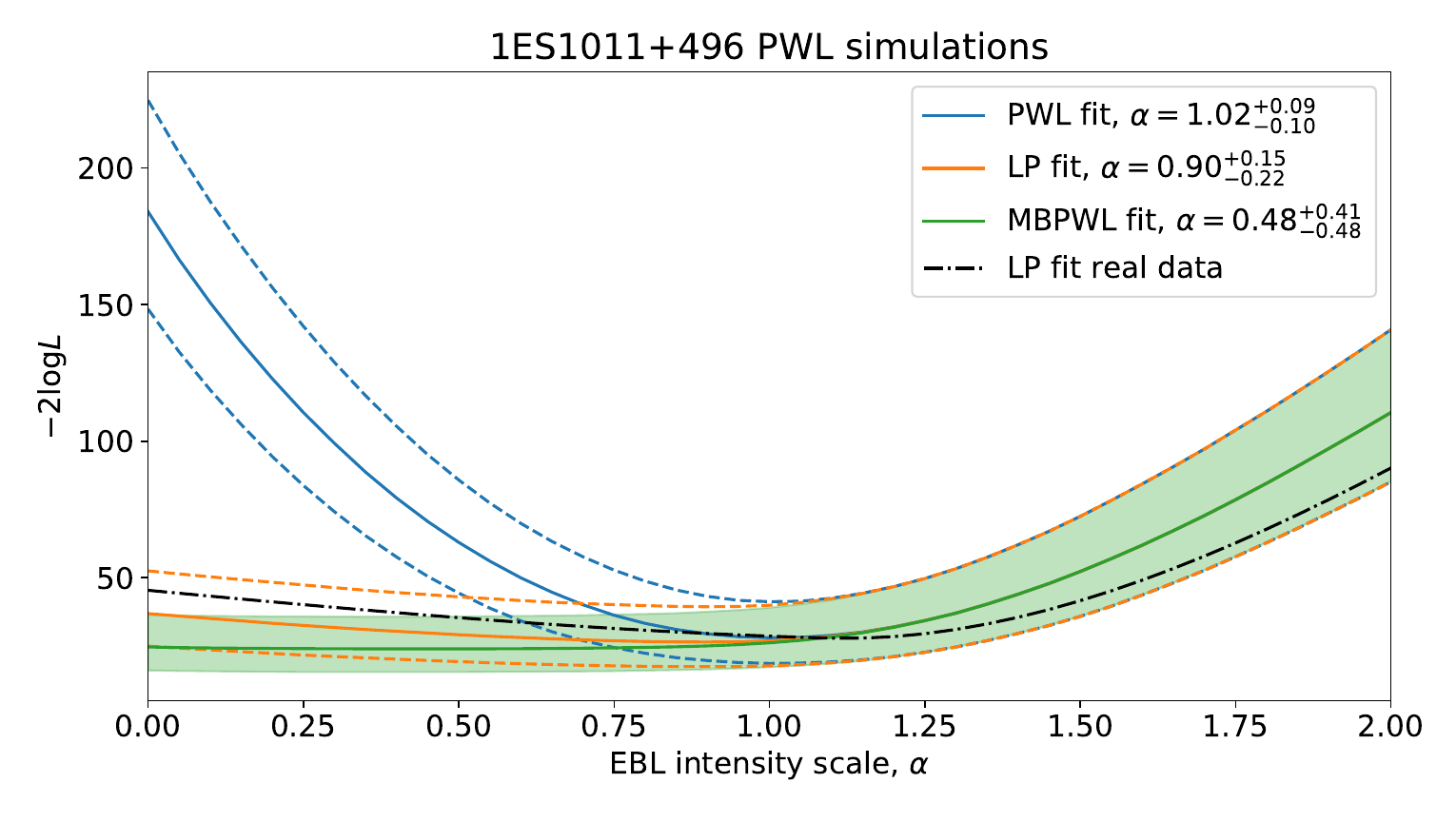}
    \caption{Median of the profile likelihood results for 1ES1011+496 simulated by a PWL and fitted with a PWL and with a LP using \protect\cite{dominguez_2011} EBL model. In dashed lines the region containing 68\% of the realizations around the median. In black dash-dotted line the profile likelihood of the real 1ES1011+496 data observed by MAGIC fitted with a LP. The green line shows the median of the results of the simulation fitted by a MBPWL and the shaded band shows the region containing 68\% of the realizations around the median.}
    \label{fig:simu_68}
\end{figure}

With those systematic error values, for 1ES1011+496, we built two profile likelihoods of $\alpha$ for each Poisson realization. One fitting a PWL and another one fitting a LP. We plotted the median of the profile likelihoods of the PWL and LP fit in \autoref{fig:simu_68} with the regions containing 68\% of the realizations around them. The uncertainties shown in the figure are the $1\sigma$ confidence intervals computed from the median likelihood profiles using Wilk's theorem: the limits of the confidence interval are defined as $\Delta (2 \textrm{log}L)$ = 1 relative to the minimum ($2 \textrm{log}L_{max}$). They can be interpreted as the "typical constraint" that would be obtained for the simulated conditions. We used the PWL fits just as a test that the code works properly - PWL is the function used to do the simulation, but using it in the analysis implies a strong and unrealistic assumption: that we know with certainty that the source spectrum has no intrinsic curvature. Indeed, the result, $\alpha = 1.02^{+ 0.09}_{- 0.10}$, features a strong lower bound, which results entirely from attributing all curvature in the observed spectrum to the EBL absorption.

\begin{figure}
    \centering
    \includegraphics[width=\linewidth]{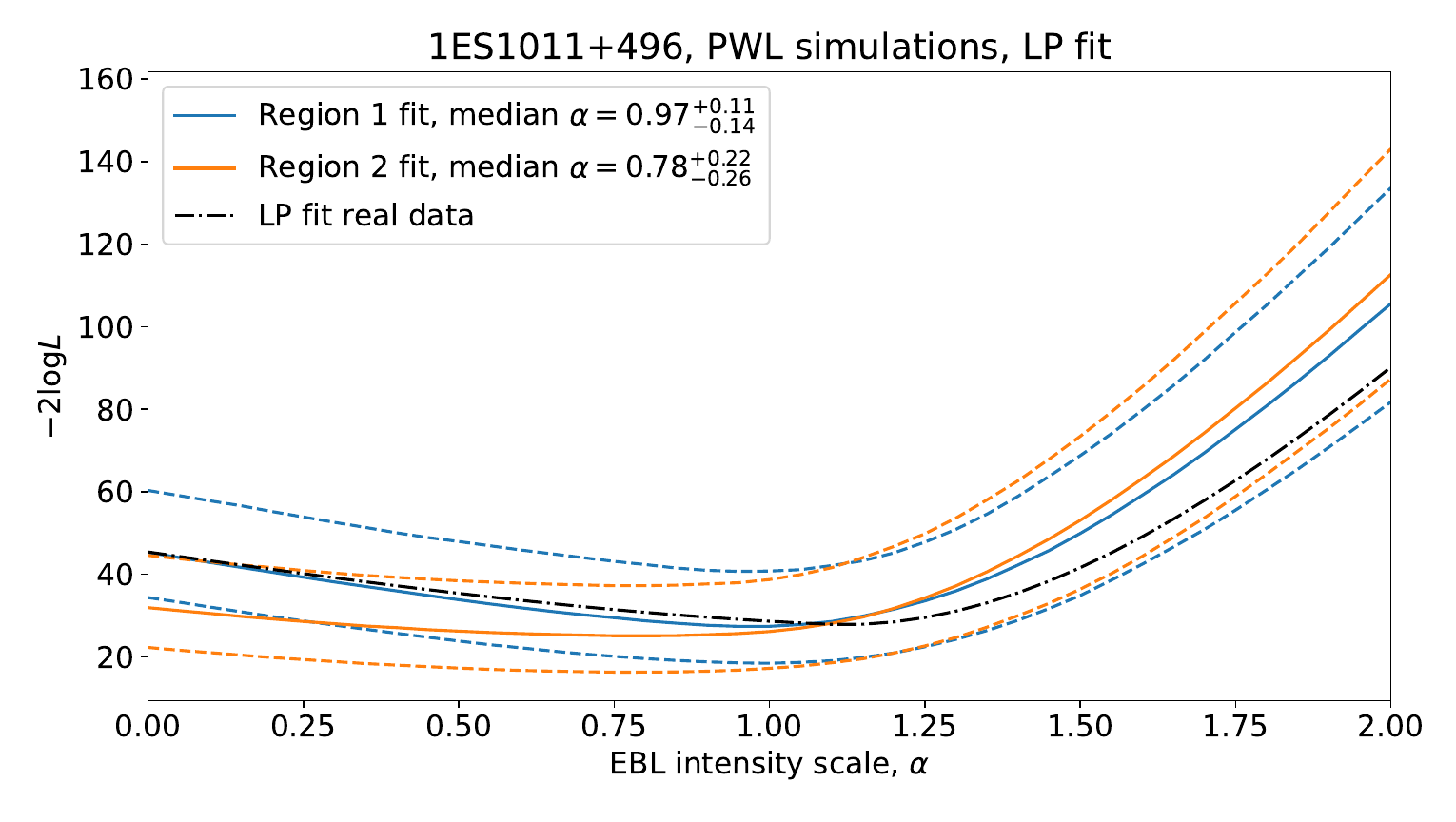}
    \caption{Median of the profile likelihood results for 1ES1011+496 simulated by a PWL and fitted by a LP. The results have been separated in 2 different regions depending on $\Delta\alpha-$. The first region has lower limits from $0.05$ to $0.2$ and the second one from $0.2$ to $0.4$.}
    \label{fig:LP_2regions}
\end{figure}

With a LP used as intrinsic spectral model (i.e. the same used for this dataset in \cite{Acciari2019} we obtain the median profile shown in orange in \autoref{fig:simu_68}). The black dot-dashed line is the actual profile from the analysis of the real data, which is contained within the $68\%$ band, but has a different shape on both sides of the minimum. A closer look at the results of the simulation shows that there are many similar cases. The shape of the curve for individual realizations can differ significantly from the {\it median} profile, depending on the bin-wise event statistics. We think this effect is particularly clear in this case, in which the simulated spectrum (PWL) is the limit case of the spectral model (concave LP): realizations in which data fluctuate towards a convex spectrum will {\it trigger} the concavity limit, which in turn affects the best-fit value of $\alpha$ and its uncertainty range. To illustrate this, in \autoref{fig:LP_2regions} we separate the realizations in two groups, depending on the value of $\Delta\alpha-$: the first group has values between $0.05$ and $0.2$, peaking at around $0.1$. The second one has $\Delta\alpha-$ between $0.2$ to $0.4$, peaking around $0.3$. The result obtained on the real data belongs to the first group.


\subsubsection{Markarian 421}

\begin{figure}
    \centering
    \includegraphics[width = \linewidth]{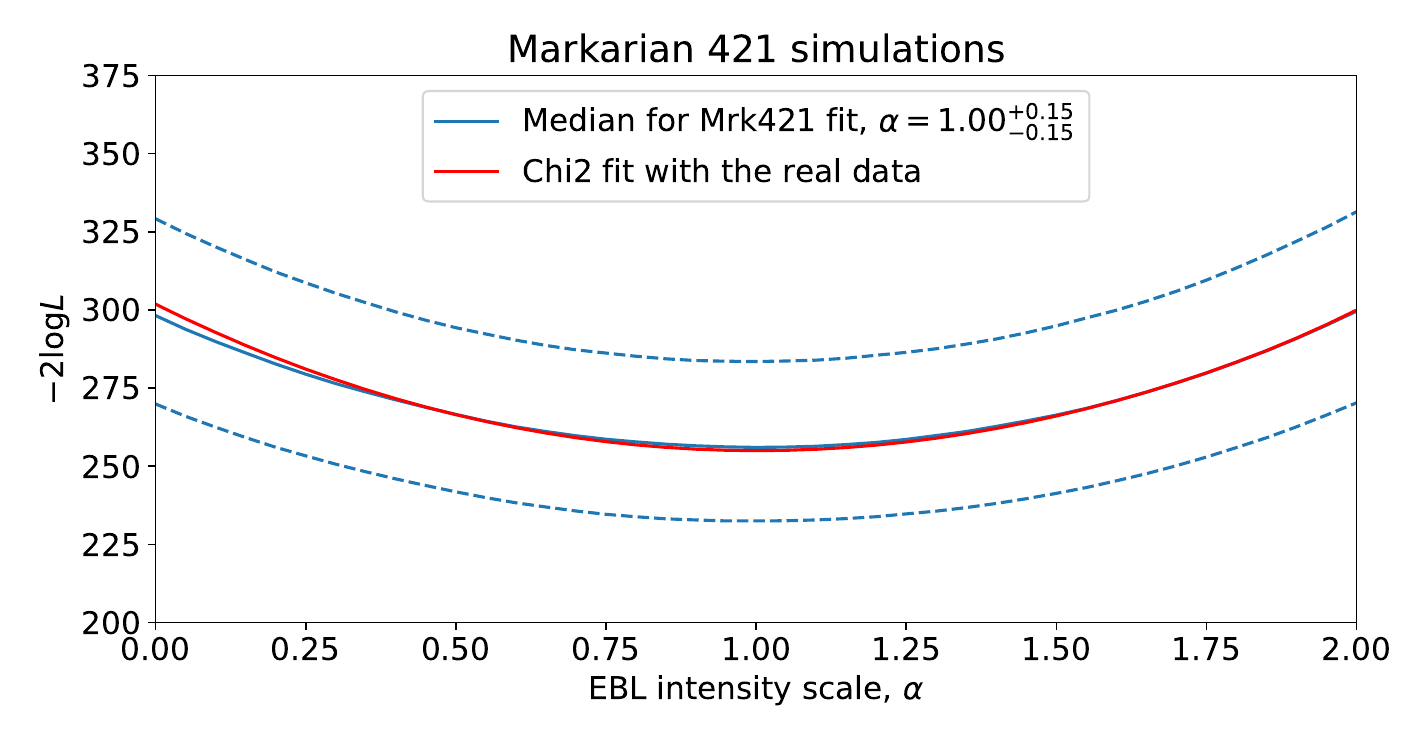}
    \caption{Results from 10,000 realizations of the simulation for the 15 Mrk421 spectra, simulated and fitted using the functions listed in \autoref{tab:Mrk_funcs} and the EBL model from \protect\cite{dominguez_2011}. In red the results of the real data fits.}
    \label{fig:Mrk_sim}
\end{figure}

For each of the observations, the analysis was performed using as intrinsic spectral model the same function used in the simulation (\autoref{tab:Mrk_funcs}). The global profile likelihood curve is obtained by adding the curves for all spectra. \autoref{fig:Mrk_sim} shows the median curve and the band around it which contains $68\%$ of the realizations, as well as the $\alpha$ uncertainty range obtained by applying Wilks' theorem to the median curve. In this case the median curve is close to the profile likelihood from the analysis of the MAGIC dataset. Unlike in the case of 1ES 1011+496, the best-fit models for the intrinsic spectra are all far from the concavity constraint, and there are no separate groups of solutions.

\begin{table*}
\caption{Functions and parameters used for Mrk421 fit and simulation. See functions in \autoref{sec:ToyMC} for the meaning of the parameters.}
\label{tab:Mrk_funcs}
\centering
\begin{tabular}{l|l|l|lllll}

Date & Obs. Time & Fit Function & $F_0$ (m$^{-2}$ s$^{-1}$& $\Gamma$ & $b$ & $d$ & $E_c$  \\
     &          &             &   TeV$^{-1}$)$\times 10^{-5}$  & & & &(TeV) \\ \hline
2013/04/10    & 0.49h     & LP           & 2.70                 & 2.28     & 0.53       &      &              \\ 
2013/04/11    & 5.49h     & SEPWL        & 6.93                 & 1.73     &            & 0.57 & 0.36         \\ 
2013/04/12    & 6.38h     & EPWL         & 5.60                 & 1.91     &            &      & 2.79         \\ 
2013/04/13a   & 1.99h     & EPWL         & 6.91                 & 1.78     &            &      & 2.31         \\ 
2013/04/13b   & 1.69h     & ELP          & 7.71                 & 1.76     & 0.12       &      & 3.35         \\ 
2013/04/13c   & 2.21h     & ELP          & 8.91                 & 1.71     & 0.10       &      & 3.53         \\ 
2013/04/14    & 6.37h     & ELP          & 5.71                 & 1.97     & 0.13       &      & 2.50         \\ 
2013/04/15a   & 1.71h     & EPWL         & 4.73                 & 1.89     &            &      & 2.79         \\ 
2013/04/15b   & 1.67h     & SEPWL        & 7.13                 & 1.68     &            & 0.52 & 1.06         \\ 
2013/04/15c   & 2.47h     & ELP          & 5.47                 & 1.76     & 0.10       &      & 8.24         \\ 
2013/04/16    & 4.13h     & ELP          & 3.02                 & 2.02     & 0.26       &      & 6.97         \\ 
2013/04/17    & 2.70h     & LP           & 1.54                 & 2.09     & 0.40       &      &              \\ 
2013/04/18    & 1.55h     & EPWL         & 1.78                 & 2.00     &            &      & 1.17         \\ 
2013/04/19    & 1.53h     & LP           & 0.92                 & 1.99     & 1.30       &      &              \\ 
2014/04/26    & 3.38h     & EPWL         & 3.88                 & 2.21     &            &      & 3.03         \\ 
\end{tabular}
\end{table*}

\subsubsection{Coverage of the results from Wilks' theorem}
With the simulation plots for 1ES1011+496 and Mrk421 we checked their $1\sigma$, $2\sigma$ and $3\sigma$ coverages. If Wilks' theorem is applicable here, those coverages should be $~68\%$, $~95\%$ and $~99.7\%$ respectively. We can see the obtained coverages in \autoref{tab:coverages}. 

\begin{table}
\caption{Coverages of the different Monte-Carlo simulations and the expected coverage. Also the coverages of the simulation with systematics in the analysis for comparison.}
\label{tab:coverages}
\begin{tabularx}{\linewidth}{p{0.32\linewidth} >{\centering\arraybackslash}p{0.15\linewidth} >{\centering\arraybackslash}p{0.15\linewidth} >{\centering\arraybackslash}p{0.15\linewidth}}

\hline
             & $1\sigma (\%)$ & $2\sigma (\%)$ & $3\sigma (\%)$ \\ \hline
Expected     & $68$          & $95$          & $99.7$       \\ 
1ES1011+496 PWL  & $49.3$        & $82.5$        & $95.8$       \\ 
1ES1011+496 LP   & $55.4$        & $86.4$        & $96.9$        \\
\mbox{1ES1011+496 MBPWL} & $37.7$      & $72.6$       & $92.2$     \\
Mrk421 (ALL) & $54.4$        & $86.8$        & $97.4$      \\ \hline
\multicolumn{1}{l}{With systematics in the analysis} & & &   \\ \hline
1ES1011+496 LP & $71.3$ & $96.45$ & $99.7$ \\
Mrk421 (ALL) & $72.6$ & $97.4$ & $99.9$ \\ \hline

\end{tabularx}
\end{table}



\begin{figure*}
    \centering
    \includegraphics[width=0.49\linewidth]{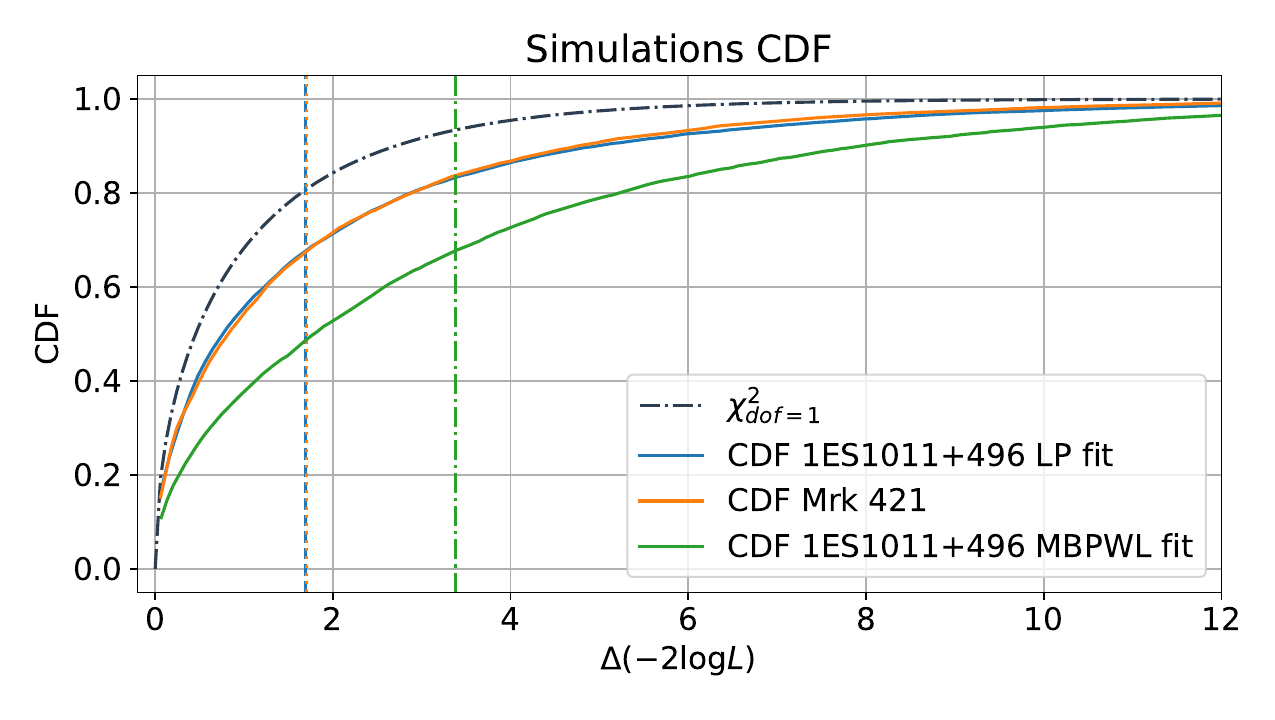}
    \hfill
    \includegraphics[width=0.49\linewidth]{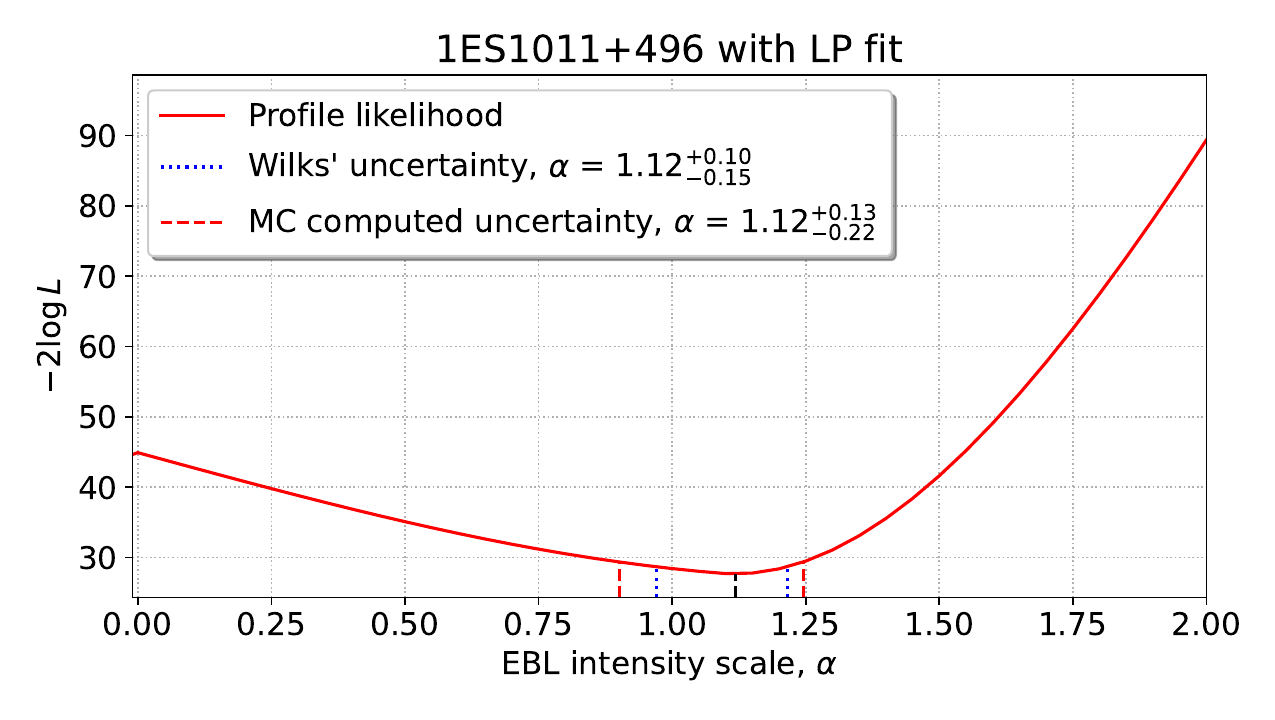}
    
    \vspace{0.3cm} 
    
    \includegraphics[width=0.49\linewidth]{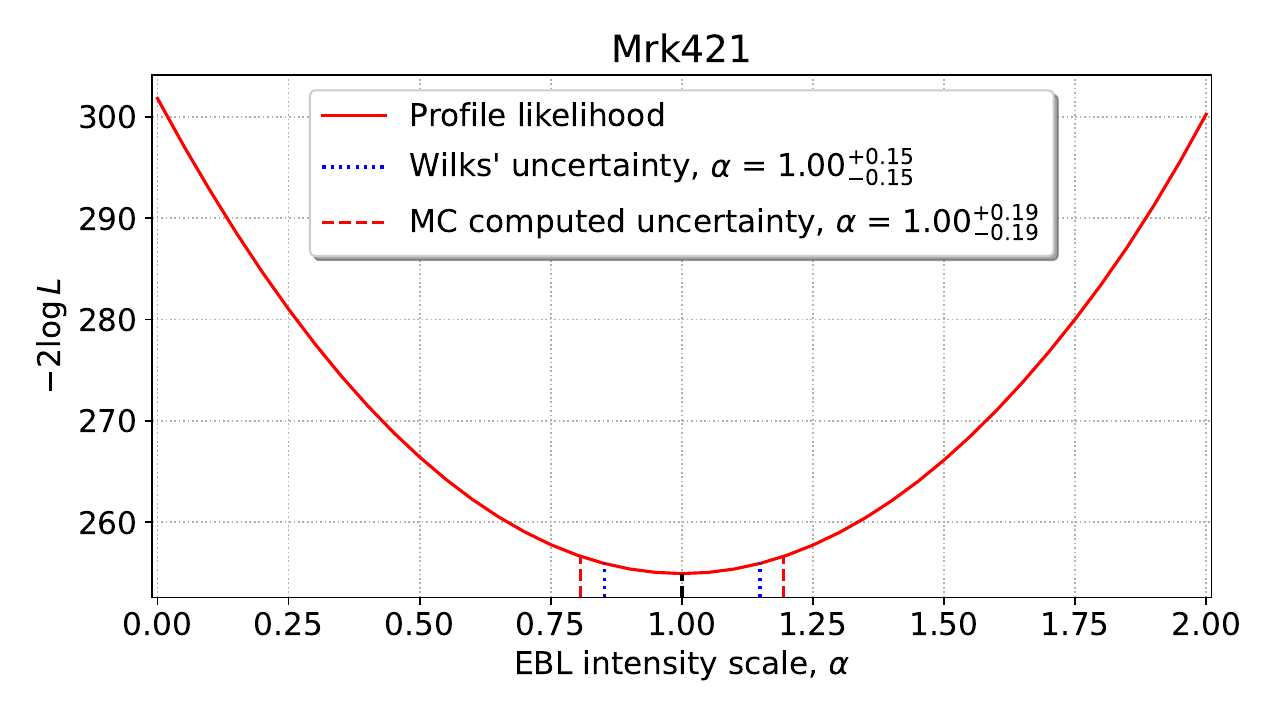}
    \hfill
    \includegraphics[width=0.49\linewidth]{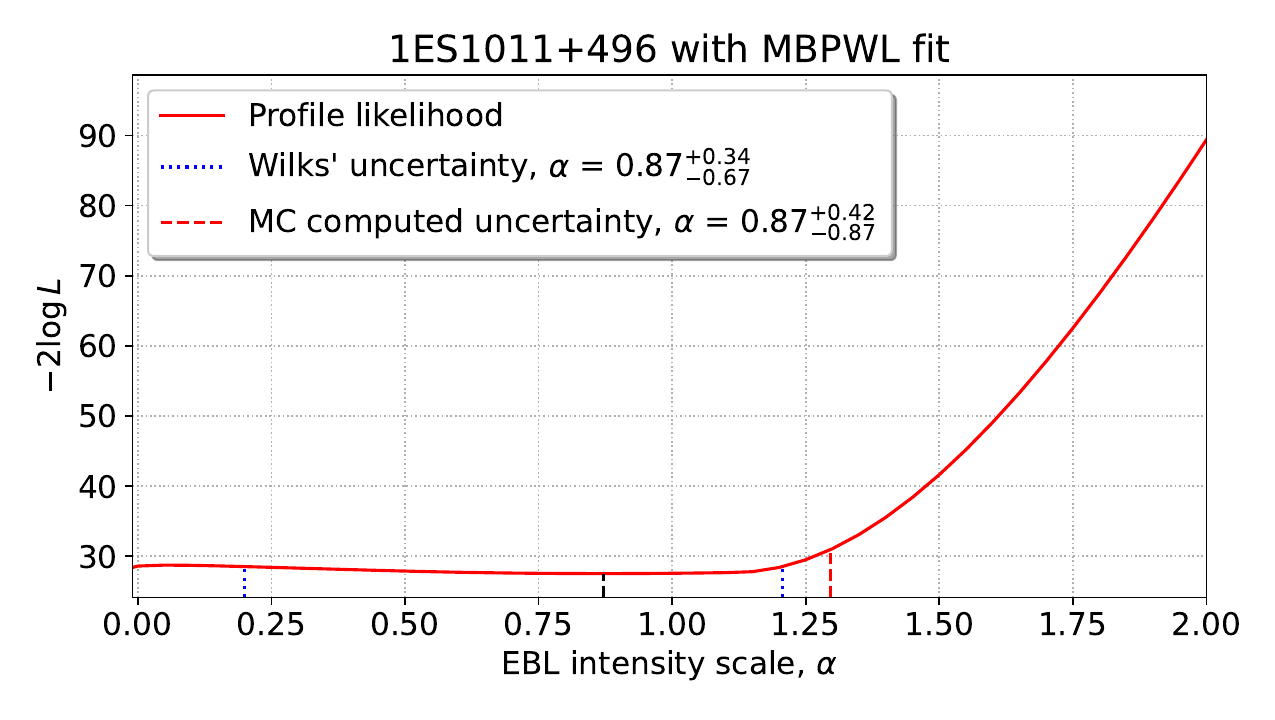}
    
    \caption{Combined simulation and real data results. 
    Top left: Cumulative Distribution Functions (CDFs) for the simulations compared to a $\chi^2$ distribution, with vertical dashed and dash-dotted indicating the $68.27\%$ thresholds. 
    Top right: Profile likelihood of the 1ES1011+496 data fitted with a log-parabola (LP).
    Bottom left: Profile likelihood of all the Mrk421 data. 
    Bottom right: Profile likelihood of the 1ES1011+496 data fitted with a MBPWL with two nodes. 
    For all profile likelihood plots, uncertainties are computed using both Wilks' theorem and the MC simulation, black dashed lines indicate the minimum positions, blue dotted lines show Wilks' uncertainties and red dashed lines MC simulation uncertainties.}
    \label{fig:combined_plots_2x2}
\end{figure*}

We can see that there is undercoverage for all cases. Now we can plot the Cumulative Distribution Function (CDF) and compare it to a $\chi^2$ with one degree of freedom. From the CDF we can obtain the value of $\Delta \textrm{log}(L)$ needed to reach the $68\%$ coverage. This is the value we will use to get the uncertainty of $\alpha$ in the analysis of the real data. In \autoref{fig:combined_plots_2x2} top left panel we can see the CDF for 1ES1011+496 simulation fitted by a LP and the CDF of the Mrk421 simulation respectively. For 1ES1011+496 fitting a LP we have found that the value of $\Delta \textrm{log}(L)$ that we need to use is $1.69$ and for Mrk421 we need to use a value of $1.70$. This value does not have to be similar for different observations as it can change with spectral shape, redshift, statistics, EBL model used. Therefore it is important to run the MC simulation for every observation.

\subsection{Application to real data}

We can now compare the uncertainties obtained via Wilks' theorem to those derived from the Monte Carlo simulations. The plots comparing these results are shown in \autoref{fig:combined_plots_2x2} top right and bottom left panels, with the corresponding data summarized in \autoref{tab:comparison_syst}: a significant increase of the 1-$\sigma$ uncertainties (relative to those from Wilks) is seen in all cases. 

\begin{figure*}
    \centering
    \includegraphics[width=0.48\linewidth]{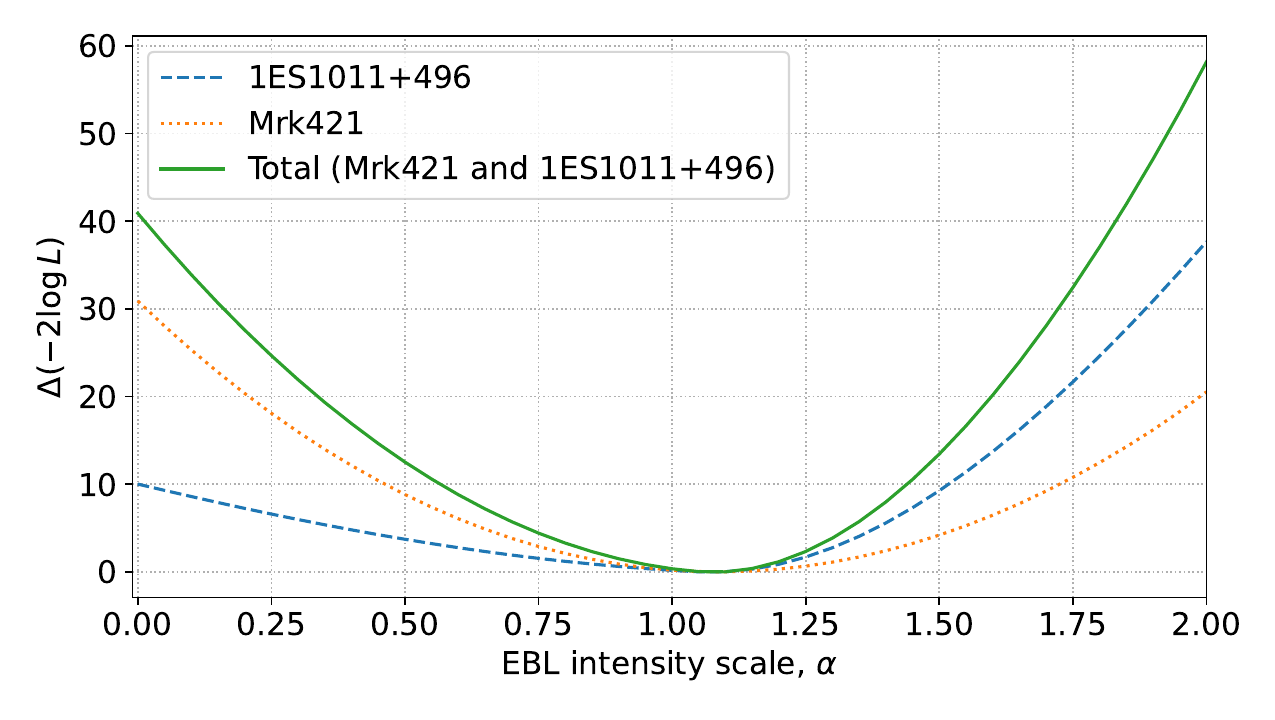}
    \hfill
    \includegraphics[width=0.48\linewidth]{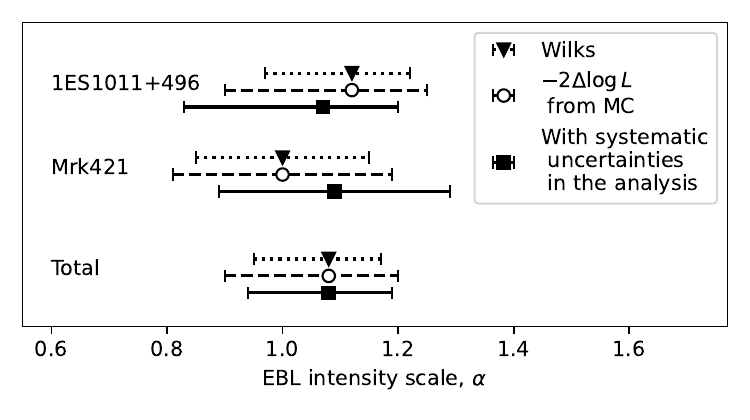}
    \caption{Left: Profile likelihood of the 1ES1011+496 and Mrk421 data, when Gaussian systematics in the effective area are introduced in the analysis as nuisance (see text). In green, the total likelihood. Right: Comparison of the $\alpha$ estimates using Wilks' theorem, the -2$\Delta \textrm{log}(L)$ from the MC simulation, and adding systematics to the analysis for 1ES1011+496 and Mrk421.}
    \label{fig:421_1011_syst_results}
\end{figure*}

Using the systematics in the analysis as explained at the end of \autoref{sec:ToyMC}, we obtain a new profile likelihood for 1ES1011+496 and Mrk421. With these new profile likelihoods (shown in the left panel of \autoref{fig:421_1011_syst_results}), we can compute the uncertainties and P-values and compare them to the ones obtained using the MC simulation.
The results are shown in \autoref{tab:comparison_syst}, \autoref{tab:comparison_syst_Pval} and \autoref{fig:421_1011_syst_results}.

\begin{table*}
\centering
\caption{Comparison between the results obtained using Wilks' theorem, the MC simulation, and adding systematics to the analysis for 1ES1011+496 (LP and MBPWL fits), Mrk421, and both sources combined. For 1ES1011+496 the LP and MBPWL fits are shown separately; for Mrk421 the fit functions used are those in \autoref{tab:Mrk_funcs}. For the combined likelihood, the 1ES1011+496 LP fit is used.}
\label{tab:comparison_syst}
\begin{tabular}{l|ccc|c}
                            & 1ES1011+496 (LP) & Mrk421  & Total & 1ES1011+496 (MBPWL) \\ \hline
Minimum                     & 1.12              & 1.00   & 1.08  & 0.87                \\
1$\sigma$ Wilks'            & +0.10-0.15        & $\pm0.15$ & +0.09-0.13 & +0.34-0.67 \\
1$\sigma$ MC simulation                & +0.13-0.22        & $\pm0.19$ & +0.12-0.18 & +0.42-0.87 \\
Minimum with systematics    & 1.07              & 1.09   & 1.08  & 1.07                \\
1$\sigma$ with systematics  & +0.13-0.24        & $\pm0.20$ & +0.11-0.14 & +0.14-0.74 \\
\end{tabular}
\end{table*}

\begin{table}
\caption{Comparison of the best-fit $-2\log L$ and P-value with and without accounting for systematics for 1ES1011+496, Mrk421, and both sources combined.}
\label{tab:comparison_syst_Pval}
\centering
\begin{tabular}{l|ccc}

                                & 1ES1011+496     & Mrk421          & Total        \\ \hline
Without systematics:               &                 &                 &              \\ 
Minimum $-2\log L$              & 27.7/16         & 254.9/220       & 283.0/237    \\
P-value                         & 0.034           & 0.053           & 0.022        \\ 
\hline
With systematics:               &                 &                 &              \\ 
Minimum $-2\log L$                      & 24.4/16         & 201.4/220       & 225.8/237    \\
P-value                         & 0.082           & 0.811           & 0.689        \\ 

\end{tabular}
\end{table}

With the total likelihood, obtained by combining both source's likelihoods, we get $\alpha = 1.08^{+0.11}_{-0.14}$ with a $\chi^2/ndof$ at the minimum of $225.8/237$ corresponding to a P-value of $0.689$.

We observe that for each of the sources the uncertainty increases to a level comparable to the MC simulation method when systematic errors are included in the analysis. However, the uncertainty does not increase as much when performing the combined analysis. The new best-fit values of $\alpha$ are somewhat closer to each other (0.02) than would be typically expected (considering the uncertainties), and this results in a deeper minimum of the joint -2$\log$L curve. Additionally, since being at different redshift, it could be that the EBL template is better for one redshift than the other, and, therefore each one, at the limit of infinite statistics, could converge at different $\alpha$ values.

Combining multiple datasets in a joint EBL analysis  is a common and powerful practice, as highlighted by e.g., \cite{Biteau_2015, Hess_2017, Biasuzzi_2019, Abeysekara_2019, Acciari2019, Desai_2019, Abramowski_2013}. We mainly focus on single-source analysis to illustrate the limitations of current methodologies, however the systematic issues are still highly relevant for multi-source studies.
\cite{Biasuzzi_2019} suggest that a 2$\sigma$ LRT threshold provides an overall sound criterion for selecting the optimal intrinsic spectral model from a nested set of functions. However, reducing the flexibility of the intrinsic spectral models on a dataset-by-dataset basis introduces a fundamental risk of overestimating the EBL intensity.
For example, for a distant source ($z\gtrsim 0.5$) where the true intrinsic spectrum has a physical curvature, a standard LRT may favor a simpler power-law model because it provides a statistically acceptable fit with fewer free parameters. This would result in an unintentional attribution of the intrinsic spectral curvature to the EBL absorption, biasing the global reconstruction. Strictly speaking, in a LRT framework, the competing models should be predefined and independent of the statistical quality of the dataset. 

\subsection{Generic concave function and concave EBL method} 
We have applied the MBPWL to the analysis of the 1ES1011+496 flare observation. The reason for this choice is that this is the single MAGIC spectrum that provided the best two-sided constraint on $\alpha$ \citep{1ES1011A, Acciari2019}. Two nodes, at $E_1 = 0.13$ TeV and $E_2 = 0.30$ TeV, are sufficient to achieve a good fit for the no EBL case (as determined with the Asimov dataset, see \autoref{subsec:mbpwl}). We can see the results of the MC simulation in \autoref{fig:simu_68} compared to the PWL and LP ones, with its coverages in \autoref{tab:coverages} and the CDF plot in \autoref{fig:combined_plots_2x2} top left panel. The $\Delta \textrm{log}(L)$ we need to use for the $68\%$ uncertainty is $3.38$. 
From \autoref{fig:simu_68} we can see how the MBPWL shares the right part of the plot, corresponding to the upper constraint, with the LP and the PWL fits. This happens because, when assuming more EBL than the simulated one, the best fit for both the MBPWL and the LP becomes a PWL. On the other hand, the MBPWL gives a much weaker lower constraint in $\alpha$ since it can fit better the overall EBL shape than the other two functions. Since this left part of the curve is so flat, the small fluctuations produced in the different Poisson realizations make some of the realizations have best fits at lower $\alpha$ values, making the median profile likelihood minimum to move towards the left. This is the reason why the coverages shown in \autoref{tab:coverages} are worse with the MBPWL than with the classical approaches. Therefore, with the MBPWL we can maintain the upper limits on $\alpha$, but we lose constraining power on the lower limit. This simply means that the better constraint from the classical approach is entirely dependent on the assumption of a specific intrinsic spectral shape (which we cannot guarantee to hold).

For the real data, we did the profile likelihood and computed its error both with Wilks' theorem and with the MC simulation. This is shown in \autoref{fig:combined_plots_2x2} bottom right panel and in \autoref{tab:comparison_syst}. It is important to mention that the lower bound obtained with the MC simulation is due to physical reasons, since the values of $\alpha < 0$ are not considered and the $\Delta \textrm{log}(L)$ value that gives the uncertainty is greater than $\Delta \textrm{log}(L)$ between the minimum and $\alpha = 0$.

\begin{figure}
    \centering
    \includegraphics[width = \linewidth]{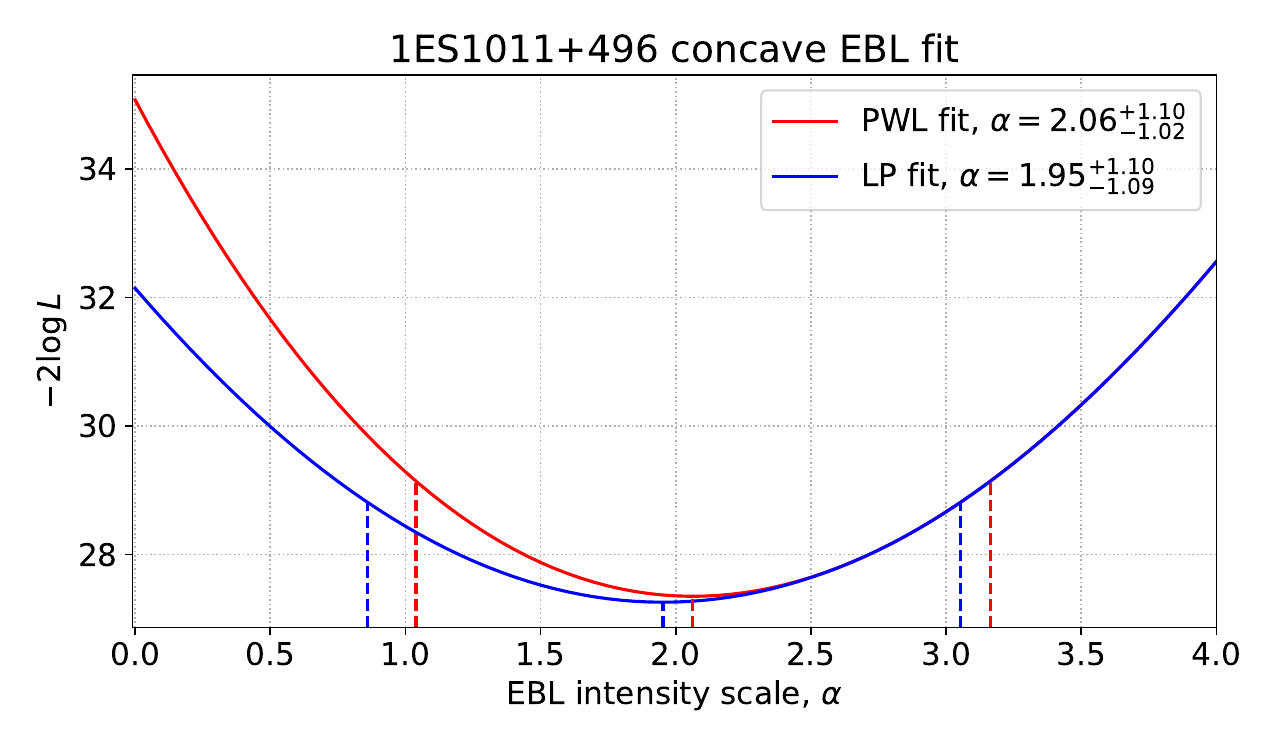}
    \caption{Profile likelihood of the 1ES1011+496 data fitted with a PWL and a LP and using the concave EBL method. Dashed lines show the minimum position and the $68\%$ uncertainty.}
    \label{fig:1ES1011_concave}
\end{figure}

For the concave EBL method, we have also used the 1ES1011+496 spectrum as a test sample. We tried both a PWL and a LP as $\frac{dF}{dE}|_{\textrm{int}}$ in \autoref{eq:tauprime}. The profile likelihood curves obtained are shown in \autoref{fig:1ES1011_concave}. As expected, the uncertainties obtained with this method are much larger than the ones from the classical approach, since now the only feature that is changed by the free parameter alpha is the EBL-induced convexity around $1$ TeV, a feature that is barely significant in the observed spectrum.

With the CDF of those simulations, we have checked that the $\Delta \textrm{log}(L)$ we need to use to get the $68\%$ uncertainty is $1.79$ for the PWL fit and $1.55$ for the LP fit.

We have also run the MC simulation with a flare similar to 1ES1011+496 in spectral shape, but 10 times brighter, to see how this method would work with more precise flux measurements. We simulated a PWL with the same photon index as in \autoref{subsec:simulation} but with $F_0 = 8.70 x 10^{-5} m^{-2} s^{-1} TeV^{-1}$. We can see a comparison of the resulting $1\sigma$ uncertainties (for $68\%$ coverage) in \autoref{tab:comparison_unc_concave}: a flare 10 times brighter than the one from 1ES1011 from February 2014 (or a similar one observed with a telescope $\simeq$10 times more sensitive than MAGIC), would still provide only a weak lower constraint on $\alpha$ ($\alpha = 0$ within $2\sigma$ of the best-fit value) if we assume the same level of systematic uncertainties (8.5$\%$). Improving the statistical uncertainties will not help unless a significant improvement in the control of systematics is achieved as well. As an example, in the table we show the case with 2$\%$ systematic uncertainties, which results in a factor $>2$ smaller uncertainty in $\alpha$. This highlights the importance of understanding and limiting the systematics in next-generation facilities, for which they will often be the dominant uncertainty.

\begin{table}
\caption{Comparison of the median best fit values (from 10000 realizations) and the uncertainties ($68\%$ coverage) of the 1ES1011+496 February 2014 flare, a hypothetical flare ten times brighter, and a flare 10 times brighter with only 2$\%$ systematics, all analyzed using the concave EBL method.}
\label{tab:comparison_unc_concave}
\begin{tabular}{l|l|l|l}
    & 1ES1011+496
    & 1ES1011+496 $\times 10$
    & 1ES1011+496 $\times 10$ \\
    & 8.5\% syst. & 8.5\% syst. & 2\% syst. \\ \hline
PWL & 1.06 + 0.94 - 0.92 & 0.97 + 0.66 - 0.53 & 1.01 + 0.25 - 0.24                 \\
LP  & 0.90 + 0.93 - 0.90 & 0.99 + 0.56 - 0.57 & 0.97 + 0.27 - 0.25                
\end{tabular}
\end{table}

\section{Summary and conclusions}\label{sec:conclusions}

The goal of this study was to critically assess the assumptions commonly adopted in EBL-constraining techniques based on extragalactic gamma-ray observations. The typical analysis makes use of a LRT in which the parameter of interest is the EBL scaling parameter $\alpha$, implemented as a global normalization of the optical depth predicted by a reference EBL template model as a function of source redshift and gamma-ray energy. We have identified the following issues in this approach: 
\begin{enumerate}
    \item Most often in the literature it has been assumed that the conditions required for Wilks' theorem to hold are fulfilled, and hence the best-fit value of $\alpha$ and its uncertainty can be obtained analytically from its profile likelihood curve. Through the use of a Monte Carlo to simulate actual observations performed with MAGIC, we have found that this assumption approximately holds (i.e. the the $1\sigma$ coverage is close to 68$\%$) only in the ideal conditions of the MC simulation, in which there are no hidden systematic errors and the models used in the analysis (for the EBL and the intrinsic source spectra) are the same ones used to perform the simulation. These conditions are however not fulfilled in the analysis of real data, as indicated by the typically low P-values of the maximum likelihood results. When systematic errors are introduced in the simulation to achieve low P-values similar to those obtained with real data, the coverage of the results falls well below the desired values.
    
    \item  The spectra of the gamma-ray sources are modeled with simple functions (of typically $\leq 4$ parameters, e.g. a log parabola) which often are by construction forced to be concave. The selection among the candidate functions is normally done using some {\it recipe} based on the goodness-of-fit of the different functions to the data sample under examination. This is in itself somewhat atypical for a Likelihood Ratio Test, in which the two competing models should in principle be defined in advance, and independently of the data. It is also a procedure which will tend to choose the simplest function which can {\it model} the observations, but in doing so it ignores the unavoidable degeneracies between intrinsic spectral features of the sources (e.g. cut-offs) and the effect of the EBL. This is best exemplified by MAGIC's observations of the 2014 1ES1011+496 flare, from which a meaningful lower constraint on $\alpha$ can be derived only under the strict assumption that the intrinsic spectrum is a power-law or a log-parabola. 
\end{enumerate}

Although our study focuses on the frequentist version of the analysis used by the MAGIC collaboration, the problems outlined above are not specific to it: any method (e.g. Bayesian approaches) which use similar assumptions would be affected as well.

\vspace{0.5cm}
We propose the following recommendations to address these issues: 
\begin{enumerate}
    \item Wilks' theorem should never be taken for granted, and MC simulations should be performed to evaluate its validity in each analysis. Particularly in the case of too low P-values, one should consider the possible existence of hidden systematic errors on top of the statistical uncertainties of the data. If Bayesian methods are used, one should also check whether the result is actually a good fit to the data, for example using the approach in \cite{DES_bayes}, to assess the possible presence of hidden systematics.
    
    Introducing in the Likelihood a simple model for the systematic uncertainties in the measured fluxes, we re-analyzed a few MAGIC datasets and obtained both larger best-fit P-values and larger uncertainties. While the specific model we adopted for the systematics is somewhat arbitrary, the result shows the importance of incorporating systematic uncertainties in the analysis. This will be even more relevant for future, better-performing instruments, for which statistical uncertainties will be significantly reduced, and likely become negligible compared to systematic ones.

    \item Regarding the problem of the selection of the intrinsic source spectral model, we submit that it may be possible to obtain meaningful constraints on $\alpha$ with the only assumption that the intrinsic source spectrum is concave. Using a "generic concave function" would hence solve the problem of function selection. Our first attempt at implementation was to use a concave-defined "multiply broken power law" as intrinsic spectral model, but this comes with its own issues (the need to define the number of breaks or "nodes", and the increasing degeneracy among the fit parameters as the number of breaks increases). When applied to the most EBL-constraining single observation recorded with MAGIC (the 2014 flare of 1ES1011+496), this approach provides no lower constraint on $\alpha$: the one from the classical analysis was indeed fully dependent on the assumption that the log parabola was the correct model for the intrinsic spectrum. As soon as we allow a more flexible modelling, the constraining power is lost because it is nearly fully degenerate with the effect of the EBL absorption. The upper constraint, on the other hand, is unaffected, because for increasing $\alpha$, the intrinsic spectrum would have to become convex in order to reproduce the observations - and this is by construction impossible for the MBPWL, just like for the concave-defined log parabola. 
    
    As an alternative to the multiply broken power law, for the determination of {\it lower} constraints on $\alpha$ we propose the "concave EBL method". The idea is to focus on the feature of the EBL imprint which would be harder to attribute to the intrinsic source spectrum: the double inflection point (or "wiggle") at around 1 TeV. For this purpose we use a version of the EBL transmissivity curve with no inflection points (hence, concave) as a fixed component of the spectral modeling, and scale freely only the remainder, i.e., the wiggle-like feature of the curve. The idea is to remove from the problem the "concave part" of the EBL imprint in the observed spectrum because it is, in principle, indistinguishable from an intrinsic feature. The lower constraint, in this approach, would be the minimum value of $\alpha$ needed to remove any significant convexity from the observed spectrum in the wiggle region. This method is obviously not suitable to set upper constraints on $\alpha$: there will be no high-energy pile-up in the intrinsic spectrum no matter how large $\alpha$ is. 
    
    To provide meaningful lower $\alpha$ constraints, however, this method (like the one using MBPWL) should be applied to spectra in which one or both of the inflection points are sampled with high significance so that the uncertainties of the flux points are much smaller than the depth of the wiggle. We are not aware of the existence of any such observation with the current generation of TeV facilities (except perhaps the observation of GRB 221009A by LHAASO), but we expect them to be common for the next generation of IACTs.

    It is worth mentioning that the upper EBL constraint on $\alpha$, which effectively caps the total EBL intensity predicted by the template model (hence limiting yet unknown contributions), is arguably the most relevant that gamma-ray astronomy can currently provide:
    lower constraints can currently be more robustly established with other methods, such as galaxy counts, and improving them with gamma-ray observations will require a better understanding of the intrinsic emission mechanisms of gamma-ray sources.

\end{enumerate}

\section*{Author contributions}
List of the main authors in alphabetical order - R. Grau: project co-leadership, paper drafting and edition, data analysis, dedicated simulations, software development, theoretical interpretation; A. Moralejo: project co-leadership, theoretical interpretation, paper drafting and edition, data analysis. The rest of the authors have contributed in one or several of the following ways: design, construction, maintenance and operation of the instrument(s); preparation and/or evaluation of the observation proposals; data acquisition, processing, calibration and/or reduction; production of analysis tools and/or related Monte Carlo simulations; discussion and approval of the contents of the draft.

\section*{ Acknowledgements} We would like to thank the Instituto de Astrof\'{\i}sica de Canarias for the excellent working conditions at the Observatorio del Roque de los Muchachos in La Palma. The financial support of the German BMFTR, MPG and HGF; the Italian INFN and INAF; the Swiss National Fund SNF; the grants PID2022-136828NB-C41, PID2022-137810NB-C22, PID2022-138172NB-C41, PID2022-138172NB-C42, PID2022-138172NB-C43, PID2022-139117NB-C41, PID2022-139117NB-C42, PID2022-139117NB-C43, PID2022-139117NB-C44, CNS2023-144504 funded by the Spanish MCIN/AEI/ 10.13039/501100011033 and "ERDF A way of making Europe"; the Indian Department of Atomic Energy; the Japanese ICRR, the University of Tokyo, JSPS, and MEXT; the Bulgarian Ministry of Education and Science, National RI Roadmap Project DO1-400/18.12.2020 and the Academy of Finland grant nr. 320045 is gratefully acknowledged. This work has also been supported by Centros de Excelencia ``Severo Ochoa'' y Unidades ``Mar\'{\i}a de Maeztu'' program of the Spanish MCIN/AEI/ 10.13039/501100011033 (CEX2019-000918-M, CEX2021-001131-S, CEX2024001442-S), by AST22\_00001\_9 with funding from NextGenerationEU funds and by the CERCA institution and grants 2021SGR00426, 2021SGR00607 and 2021SGR00773 of the Generalitat de Catalunya; by the Croatian Science Foundation (HrZZ) Project IP-2022-10-4595 and by the University of Rijeka Project uniri-mzi-25-3 funded by the European Union - NextGenerationEU; by the Deutsche Forschungsgemeinschaft (SFB1491) and by the Lamarr-Institute for Machine Learning and Artificial Intelligence; by the Polish Ministry of Science and Higher Education grant No. 2025/WK/04; by the European Union (ERC, MicroStars, 101076533); and by the Brazilian MCTIC, the CNPq Productivity Grant 309053/2022-6 and FAPERJ Grants E-26/200.532/2023 and E-26/211.342/2021. The research leading to these results has received funding from the ESF under the program Ayudas predoctorales of the Ministerio de Ciencia e Innovación  PRE2020-093561.

\section*{Data Availability}

 The analysis software used to obtain the results presented in this article is available at \url{https://github.com/R-Grau/EBLpy}. The data underlying this article will be shared on reasonable request to the corresponding authors.



\bibliographystyle{mnras}
\bibliography{bibliography} 








\clearpage
\section*{Affiliations}
$^{1}$ {Japanese MAGIC Group: Department of Physics, Tokai University, Hiratsuka, 259-1292 Kanagawa, Japan}\\
$^{2}$ {Japanese MAGIC Group: Department of Physics, Kyoto University, 606-8502 Kyoto, Japan} \\
$^{3}$ {ETH Z\"urich, CH-8093 Z\"urich, Switzerland} \\
$^{4}$ {Universit\`a di Siena and INFN Pisa, I-53100 Siena, Italy} \\
$^{5}$ {Institut de F\'isica d'Altes Energies (IFAE), The Barcelona Institute of Science and Technology (BIST), E-08193 Bellaterra (Barcelona), Spain} \\
$^{6}$ {Universitat de Barcelona, ICCUB, IEEC-UB, E-08028 Barcelona, Spain} \\
$^{7}$ {Instituto de Astrof\'isica de Andaluc\'ia-CSIC, Glorieta de la Astronom\'ia s/n, 18008, Granada, Spain} \\
$^{8}$ {Universit\`a di Padova and INFN, I-35131 Padova, Italy} \\
$^{9}$ {National Institute for Astrophysics (INAF), I-00136 Rome, Italy} \\
$^{10}$ {Universit\`a di Udine and INFN Trieste, I-33100 Udine, Italy} \\
$^{11}$ {Max-Planck-Institut f\"ur Physik, D-85748 Garching, Germany} \\
$^{12}$ {Instituto de Astrof\'isica de Canarias and Dpto. de  Astrof\'isica, Universidad de La Laguna, E-38200, La Laguna, Tenerife, Spain} \\
$^{13}$ {Croatian MAGIC Group: University of Zagreb Faculty of Electrical Engineering and Computing, Unska 3, 10000 Zagreb, Croatia} \\
$^{14}$ {Saha Institute of Nuclear Physics, A CI of Homi Bhabha National Institute, Kolkata 700064, West Bengal, India} \\
$^{15}$ {Centro Brasileiro de Pesquisas F\'isicas (CBPF), 22290-180 URCA, Rio de Janeiro (RJ), Brazil} \\
$^{16}$ {IPARCOS Institute and EMFTEL Department, Universidad Complutense de Madrid, E-28040 Madrid, Spain} \\
$^{17}$ {Japanese MAGIC Group: Institute for Cosmic Ray Research (ICRR), The University of Tokyo, Kashiwa, 277-8582 Chiba, Japan} \\
$^{18}$ {University of Lodz, Faculty of Physics and Applied Informatics, Department of Astrophysics, 90-236 Lodz, Poland} \\
$^{19}$ {Centro de Investigaciones Energ\'eticas, Medioambientales y Tecnol\'ogicas, E-28040 Madrid, Spain} \\
$^{20}$ {Departament de F\'isica, and CERES-IEEC, Universitat Aut\`onoma de Barcelona, E-08193 Bellaterra, Spain} \\
$^{21}$ {Universit\`a di Pisa and INFN Pisa, I-56126 Pisa, Italy} \\
$^{22}$ {INFN MAGIC Group: INFN Sezione di Bari and Dipartimento Interateneo di Fisica dell'Universit\`a e del Politecnico di Bari, I-70125 Bari, Italy} \\
$^{23}$ {Department for Physics and Technology, University of Bergen, Norway} \\
$^{24}$ {INFN MAGIC Group: INFN Sezione di Torino and Universit\`a degli Studi di Torino, I-10125 Torino, Italy} \\
$^{25}$ {Croatian MAGIC Group: University of Rijeka, Faculty of Physics, 51000 Rijeka, Croatia} \\
$^{26}$ {Universit\"at W\"urzburg, D-97074 W\"urzburg, Germany} \\
$^{27}$ {Technische Universit\"at Dortmund, D-44221 Dortmund, Germany} \\
$^{28}$ {Japanese MAGIC Group: Physics Program, Graduate School of Advanced Science and Engineering, Hiroshima University, 739-8526 Hiroshima, Japan} \\
$^{29}$ {Armenian MAGIC Group: ICRANet-Armenia, 0019 Yerevan, Armenia} \\
$^{30}$ {Croatian MAGIC Group: Josip Juraj Strossmayer University of Osijek, Department of Physics, 31000 Osijek, Croatia} \\
$^{31}$ {Finnish MAGIC Group: Finnish Centre for Astronomy with ESO, Department of Physics and Astronomy, University of Turku, FI-20014 Turku, Finland} \\
$^{32}$ {University of Geneva, Chemin d'Ecogia 16, CH-1290 Versoix, Switzerland} \\
$^{33}$ {Inst. for Nucl. Research and Nucl. Energy, Bulgarian Academy of Sciences, BG-1784 Sofia, Bulgaria} \\
$^{34}$ {INFN MAGIC Group: INFN Sezione di Catania and Dipartimento di Fisica e Astronomia, University of Catania, I-95123 Catania, Italy} \\
$^{35}$ {Finnish MAGIC Group: Space Physics and Astronomy Research Unit, University of Oulu, FI-90014 Oulu, Finland} \\
$^{36}$ {Japanese MAGIC Group: Institute for Space-Earth Environmental Research and Kobayashi-Maskawa Institute for the Origin of Particles and the Universe, Nagoya University, 464-6801 Nagoya, Japan} \\
$^{37}$ {INFN MAGIC Group: INFN Roma Tor Vergata, I-00133 Roma, Italy} \\
$^{38}$ {also at International Center for Relativistic Astrophysics (ICRA), Rome, Italy} \\
$^{39}$ {also at Como Lake centre for AstroPhysics (CLAP), DiSAT, Universit\`a dell?Insubria, via Valleggio 11, 22100 Como, Italy.} \\
$^{40}$ {also at Port d'Informaci\'o Cient\'ifica (PIC), E-08193 Bellaterra (Barcelona), Spain} \\
$^{41}$ {also at Department of Physics, University of Oslo, Norway} \\
$^{42}$ {also at Dipartimento di Fisica, Universit\`a di Trieste, I-34127 Trieste, Italy} \\
$^{43}$ {also at Dipartimento di Fisica, Universit\`a di Roma Tor Vergata, Via della Ricerca Scientifica, 1, Roma I-00133, Italy} \\
$^{44}$ {also at INAF Padova} \\

\bsp	
\label{lastpage}

\end{document}